\documentclass[aps, prb, longbibliography, twocolumn]{revtex4-2}
\usepackage[pdftex]{graphicx}
\usepackage{color}
\usepackage{amsmath,amsthm,amssymb,mathtools}
\usepackage[colorlinks=true,allcolors=blue]{hyperref}

\begin{document}
\title{Graphene-hBN interlayer interactions from quantum Monte Carlo}

\author{Kittithat Krongchon}
\affiliation{Department of Physics, Institute for Condensed Matter Theory, University of Illinois at Urbana-Champaign}
\author{Tawfiqur Rakib}
\affiliation{Department of Mechanical Science and Engineering, University of Illinois at Urbana-Champaign}
\author{Daniel Palmer}
\affiliation{Department of Materials Science and Engineering, University of Illinois at Urbana-Champaign}
\author{Elif Ertekin}
\affiliation{Department of Mechanical Science and Engineering, University of Illinois at Urbana-Champaign}
\affiliation{Materials Research Laboratory, University of Illinois at Urbana-Champaign}
\author{Harley T. Johnson}
\affiliation{Department of Mechanical Science and Engineering, University of Illinois at Urbana-Champaign}
\affiliation{Department of Materials Science and Engineering, University of Illinois at Urbana-Champaign}
\author{Lucas K. Wagner}
\affiliation{Department of Physics, Institute for Condensed Matter Theory, University of Illinois at Urbana-Champaign}

\date{\today}

\begin{abstract}
The interaction between graphene and hexagonal boron nitride (hBN) plays a pivotal role in determining the electronic and structural properties of graphene-based devices. In this work, we employ quantum Monte Carlo (QMC) to study the interlayer interactions and stacking-fault energy (SFE) between graphene and hBN. We generated QMC energies for several rigid bilayer stacking configurations and fitted these data to the Kolmogorov-Crespi type interlayer potential (ILP) model. Our QMC-derived potential offers a more reliable alternative to conventional density functional theory methods, which are prone to errors in predicting properties in van der Waals materials. This study enables highly accurate predictions of structural and electronic properties in graphene/hBN heterostructures. The resulting ILP-QMC potential is made available for further use in simulating complex systems, such as twisted bilayer graphene (TBG) on hBN.
\end{abstract}

\maketitle
\section{Introduction}

Hexagonal boron nitride (hBN) is a layered material that has emerged as an essential substrate for studying graphene due to its mechanical support, smooth surface, and ability to enhance carrier mobilities in graphene devices~\cite{bao2009controlled,morgenstern2011scanning,andrei2012electronic,xue2011scanning}.
Each hBN layer can be exfoliated from bulk boron nitride, having the hexagonal lattice structure that contains boron and nitrogen atoms occupying the two sublattice sites~\cite{novoselov2005two,gorbachev2010hunting}.
The sublattice asymmetry results in a wide band gap of approximately 6 eV~\cite{zunger1976optical,watanabe2004direct,yankowitz2019van,elias2019direct}.
This insulating property of hBN minimizes electronic interference with the graphene layer, thereby allowing graphene to retain its high carrier mobility and Dirac-like electronic structure when placed on an hBN substrate, as compared to more electrically active substrates, such as black phosphorus~\cite{qiu2017environmental}.

Despite being an insulator, hBN alters the electronic structure of the graphene system~\cite{dean2010boron,hunt2013massive}.
This characteristic has led to its widespread use in heterostructure applications~\cite{geim2013van}, such as optoelectronics~\cite{aggoune2017enhanced,withers2015light}, spintronics~\cite{kamalakar2014spintronics}, sensors~\cite{he2014graphene}, DNA sequencing~\cite{shukla2017prospects}, and high-temperature devices~\cite{vsivskins2019high}.
Furthermore, the moir\'{e} pattern that forms due to the small lattice mismatch~\cite{song2013electron} between graphene and hBN gives rise to unique phenomena, such as emergence of secondary Dirac points~\cite{yankowitz2012emergence,ponomarenko2013cloning}, Hofstadter's butterfly~\cite{dean2013hofstadter}, the commensurate-incommensurate transition~\cite{woods2014commensurate}, and tunable flat bands~\cite{chen2020tunable}.

In recent years, significant attention has been directed toward twisted bilayer graphene (TBG) since the discovery of correlated phases in TBG~\cite{carr2020electronic}, such as superconductivity and insulating states~\cite{cao2018unconventional,cao2018correlated,padhi2018doped,padhi2019pressure}.
These phases are attributed to the flattening of bands near the Fermi level, which arises from moir\'{e} superlattices formed by the small twist angle between two graphene layers~\cite{bistritzer2011moire,koshino2018maximally,po2018origin,yankowitz2019tuning}.
A previous study using atomistic and tight-binding calculations reports that the band structure of TBG is sensitive to the twist angle relative to the hBN substrate, where this dependence can open a band gap and break the layer degeneracy~\cite{long2022atomistic}.
Since hBN is commonly used as a substrate to in graphene experiments, the interactions between graphene and hBN need to be accurately modeled to understand the behavior of more complex systems.

In layered materials, van der Waals (vdW) interactions play a pivotal role in determining their behavior~\cite{reina2009large,rong1993electronic,zhou2006low,peres2006electronic,shima1993electronic,nguyen2014excitation,can2021high,peng2022observation,peng2024observation}.
These interlayer interactions influence the stacking-fault energy (SFE), which in turn determines the stability of different stacking configurations in bilayer and multi-layer systems~\cite{zhou2015van,zhang2024impact}.
Thus, an accurate assessment of interlayer interactions and the SFE in graphene/hBN systems is particularly important in studying moir\'{e} patterns and their effects on material behavior.

One major challenge in modeling graphene/hBN systems lies in the correlated nature of vdW interactions. 
Conventional methods such as density functional theory (DFT) suffer from large errors due to approximate electron correlations through functionals~\cite{dubecky2016noncovalent,krongchon2017accurate,williams2020direct}.
Previous studies have employed DFT or DFT-trained potentials to model graphene and hBN systems~\cite{giovannetti2007substrate,correa2014optical,sevilla2021graphene,long2022atomistic}.
However, the discrepancy across different DFT functionals and vdW correction schemes lead to unreliable results~\cite{mostaani2015quantum}.
Different vdW models can result in a 13\% error in the binding energy, 50\% error in the corrugation and a 25\% error in flat bandwidth at the magic twist angle of $0.99^{\circ}$~\cite{krongchon2023registry}.
While the so-called many-body dispersion (MBD) correction method~\cite{ambrosetti2014long} seems promising in predicting the corrugated structure and the flat bandwidth of TBG~\cite{krongchon2023registry}, this agreement has not been established across different material types, such as graphene/hBN heterobilayer.
In fact, as we will show later, the MBD-fitted potential leads to a large error in the graphene/hBN interlayer interaction and SFE.
As such, a more reliable method is required to study vdW materials.

Quantum Monte Carlo (QMC) is a powerful computation technique, which explicitly includes electron correlations in the many-body wave function, allowing vdW interactions to emerge naturally without the need for semi-empirical corrections. 
QMC has demonstrated success across a wide range of condensed matter systems, including vdW materials~\cite{gurtubay2007dissociation,shulenburger2015nature,wu2016hexagonal,kadioglu2018diffusion,ahn2021metastable,staros2022combined,wines2023quantum}.
In our previous work~\cite{krongchon2023registry}, we used QMC to assess the sensitivity of flat bands to corrugation in free-standing TBG, revealing the impact of vdW interactions on electronic properties.
Although we have provided QMC-level accuracy for the vdW interactions between graphene bilayers, the QMC data for graphene-hBN interactions and SFE is absent.

In this study, we address this knowledge gap by focusing on the interlayer interactions between graphene and hBN.
We generated QMC energies as a function of displacement for four rigid bilayer registries: AB, SP, BA, AA, as diagrammed in Fig.~\ref{fig:stack}.
The QMC energies were then used to fit the interlayer potential (ILP), a Kolmogorov--Crespi-type potential~\cite{kolmogorov2005registry,ouyang2018nanoserpents}.
The fitted potential, suitable for simulations in \texttt{LAMMPS}~\cite{ouyang2018nanoserpents,thompson2022lammps}, offers a refined tool for simulating the geometry and electronic properties of TBG on hBN, while eliminating the uncertainty from using semi-empirical methods. The QMC-fitted potential file and energy data are made available in Ref.~\cite{krongchon2025data}.

\subsection{Quantum Monte Carlo}

\begin{figure}
    \centering
    \includegraphics{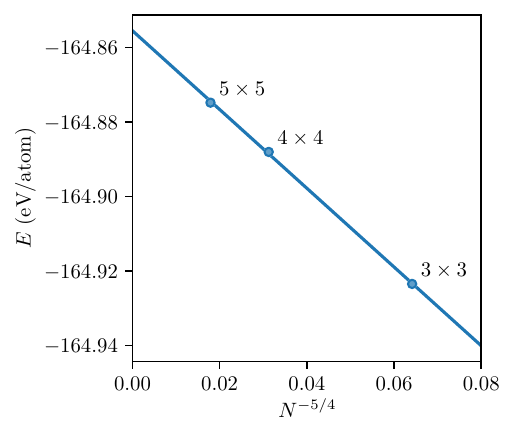}
    \caption{
    QMC energy vs. $N^{-5/4}$, where $N$ is the number of primitive cells in a simulation cell.
    The extrapolated energy $E(N \rightarrow \infty)$ is calculated by fitting Eq.~\ref{eq:hbn_qmc_extrap} using $N = 3 \times 3, 4 \times 4$, and $5 \times 5$.
    The error bar of the extrapolated energy is obtained from the bootstrapping technique.
    }
    \label{fig:hbn_extrap}
\end{figure}

\begin{figure*}
    \centering
    \begin{tabular}{c}
        Graphene/Graphene \\\\
        \begin{minipage}[t]{0.24\textwidth}
            \centering
            AB $(s = 0)$ \\
            \includegraphics[height=0.9in]{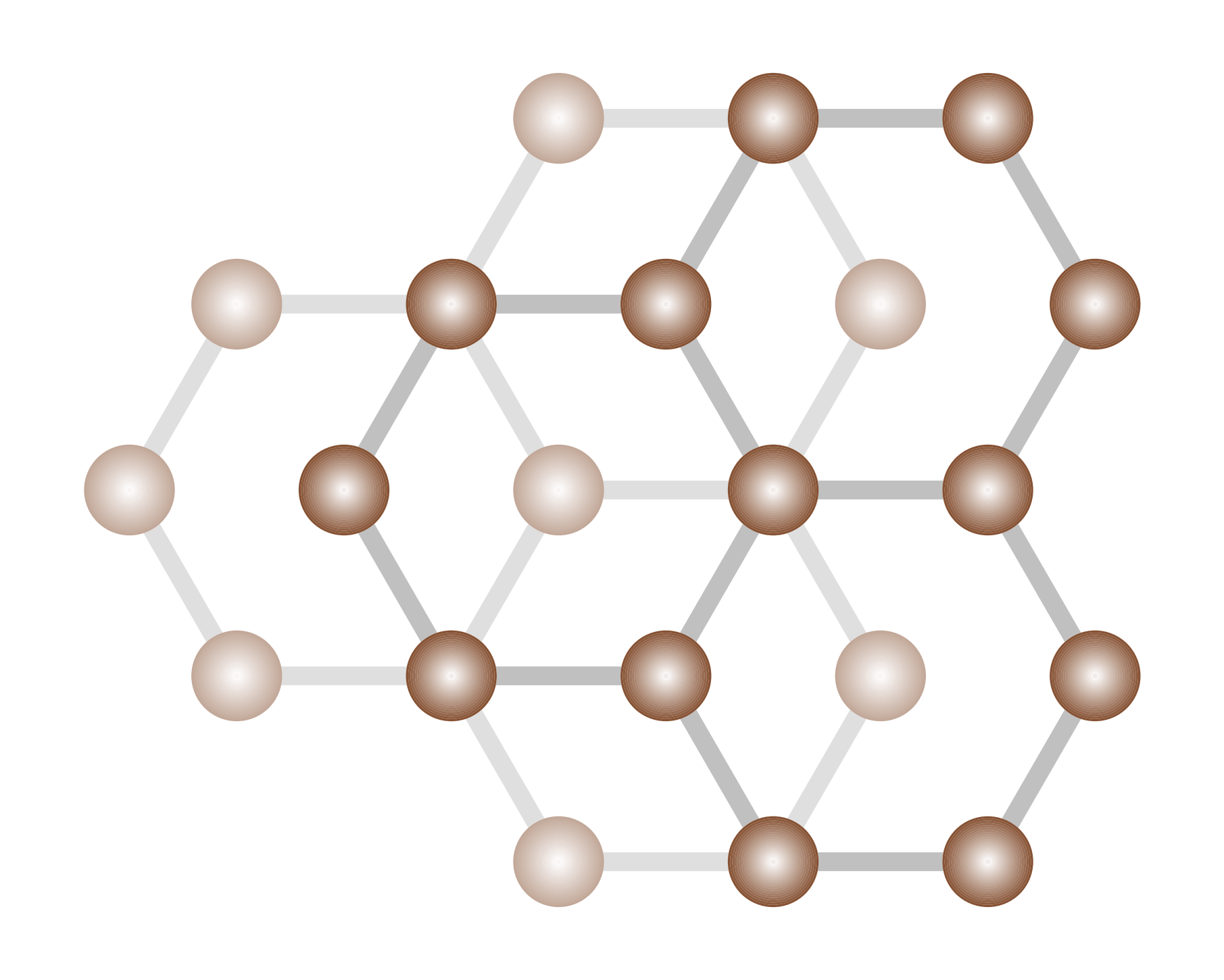}
        \end{minipage}
        \begin{minipage}[t]{0.24\textwidth}
            \centering
            SP $(s = 1/6)$ \\
            \includegraphics[height=0.9in]{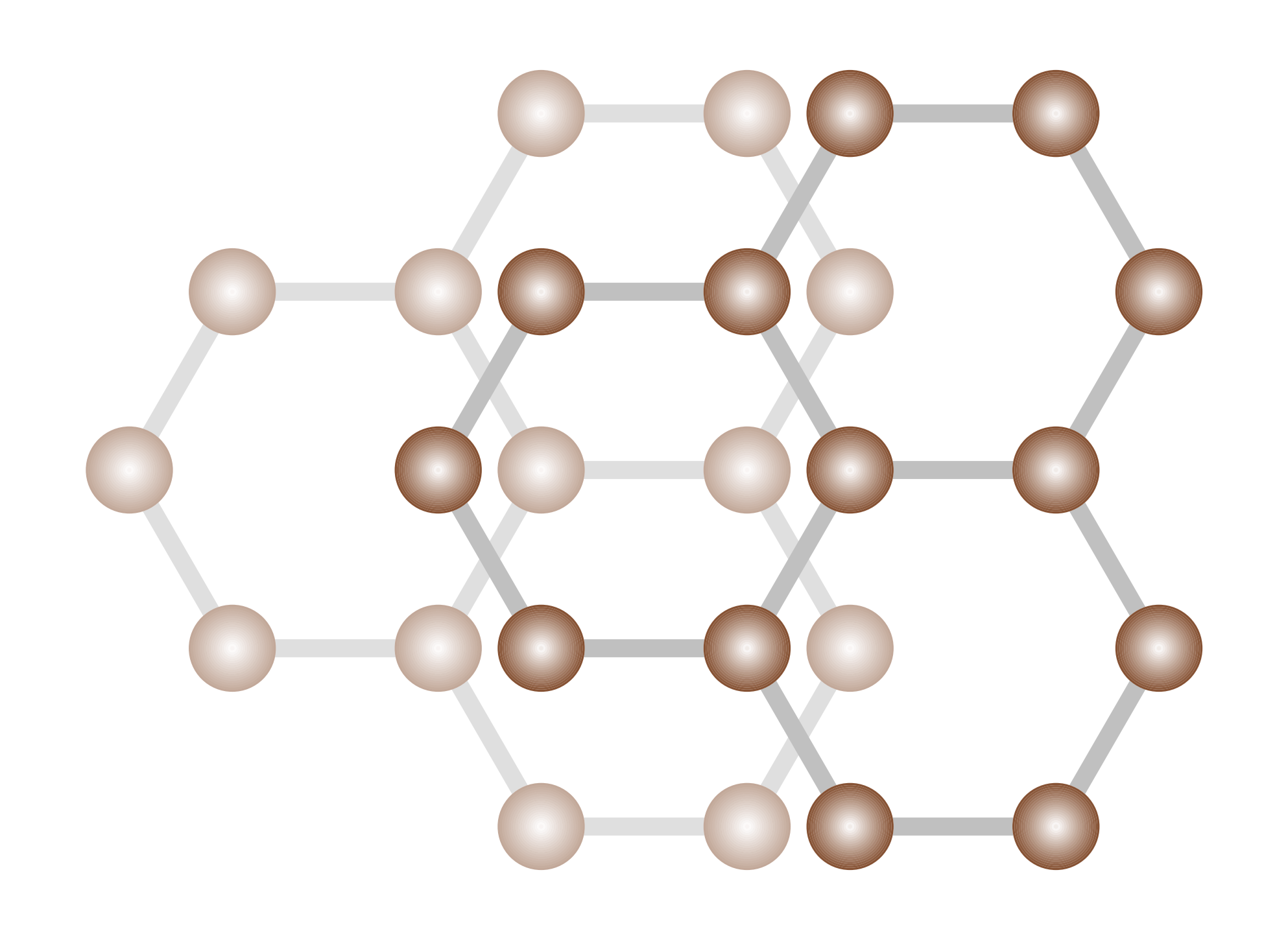}
        \end{minipage}
        \begin{minipage}[t]{0.24\textwidth}
            \centering
            Mid $(s = 1/2)$ \\
            \includegraphics[height=0.9in]{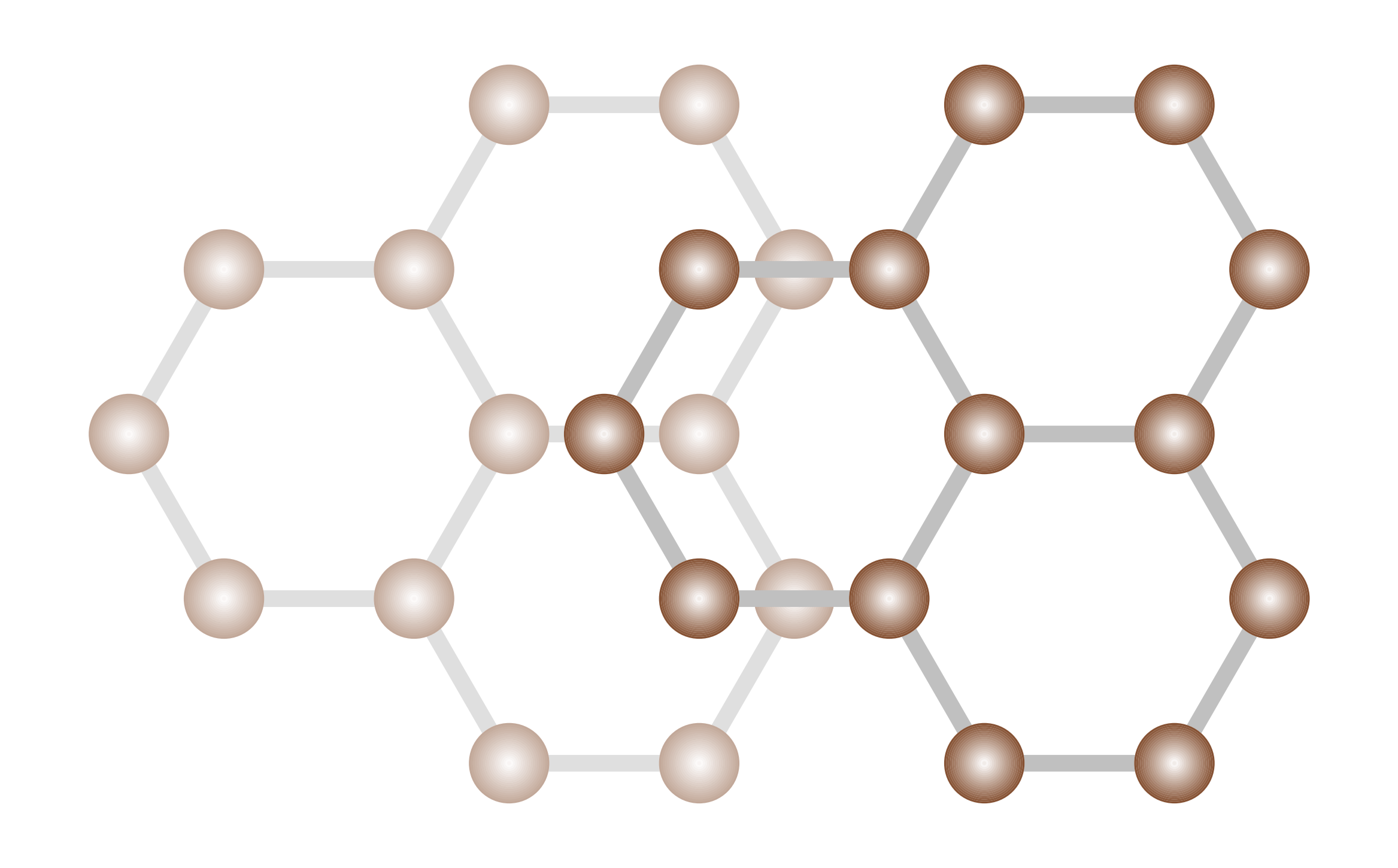}
        \end{minipage}
        \begin{minipage}[t]{0.24\textwidth}
            \centering
            AA $(s = 2/3)$ \\
            \includegraphics[height=0.9in]{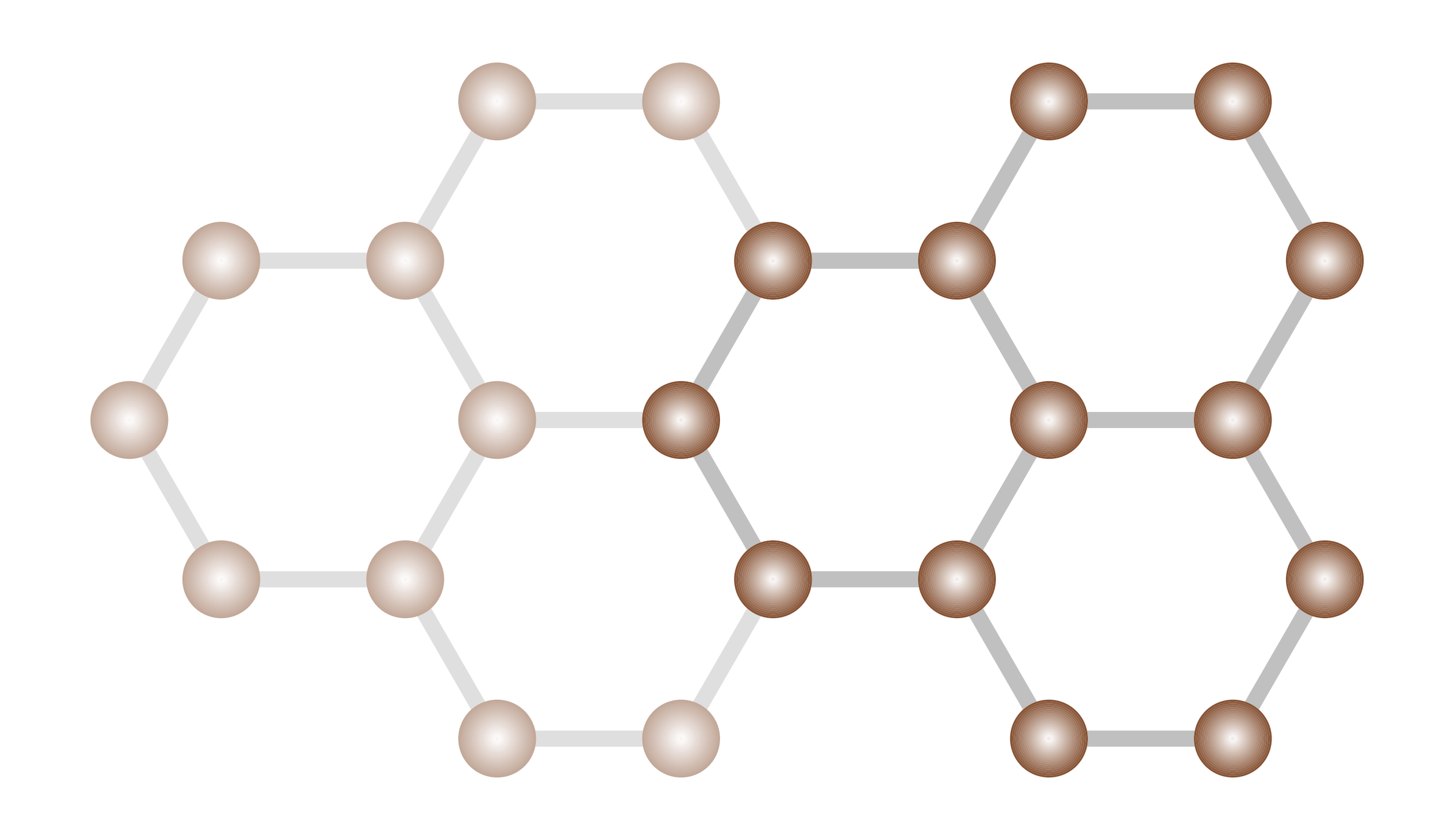}
        \end{minipage}
        \\\\
        Graphene/hBN \\\\
        \begin{minipage}[t]{0.24\textwidth}
            \centering
            AB $(s = 0)$ \\
            \includegraphics[height=0.9in]{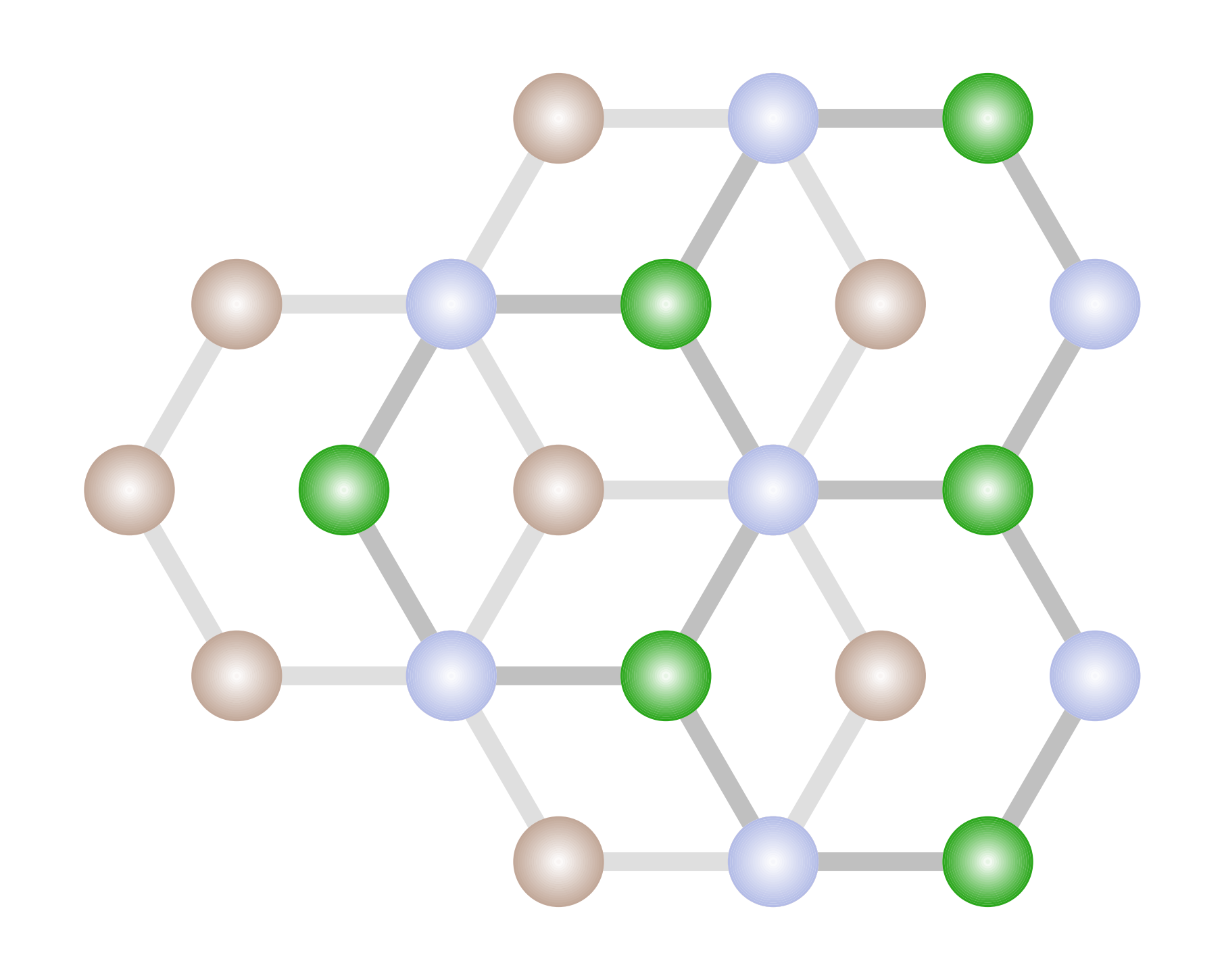}
        \end{minipage}
        \begin{minipage}[t]{0.24\textwidth}
            \centering
            SP $(s = 1/6)$ \\
            \includegraphics[height=0.9in]{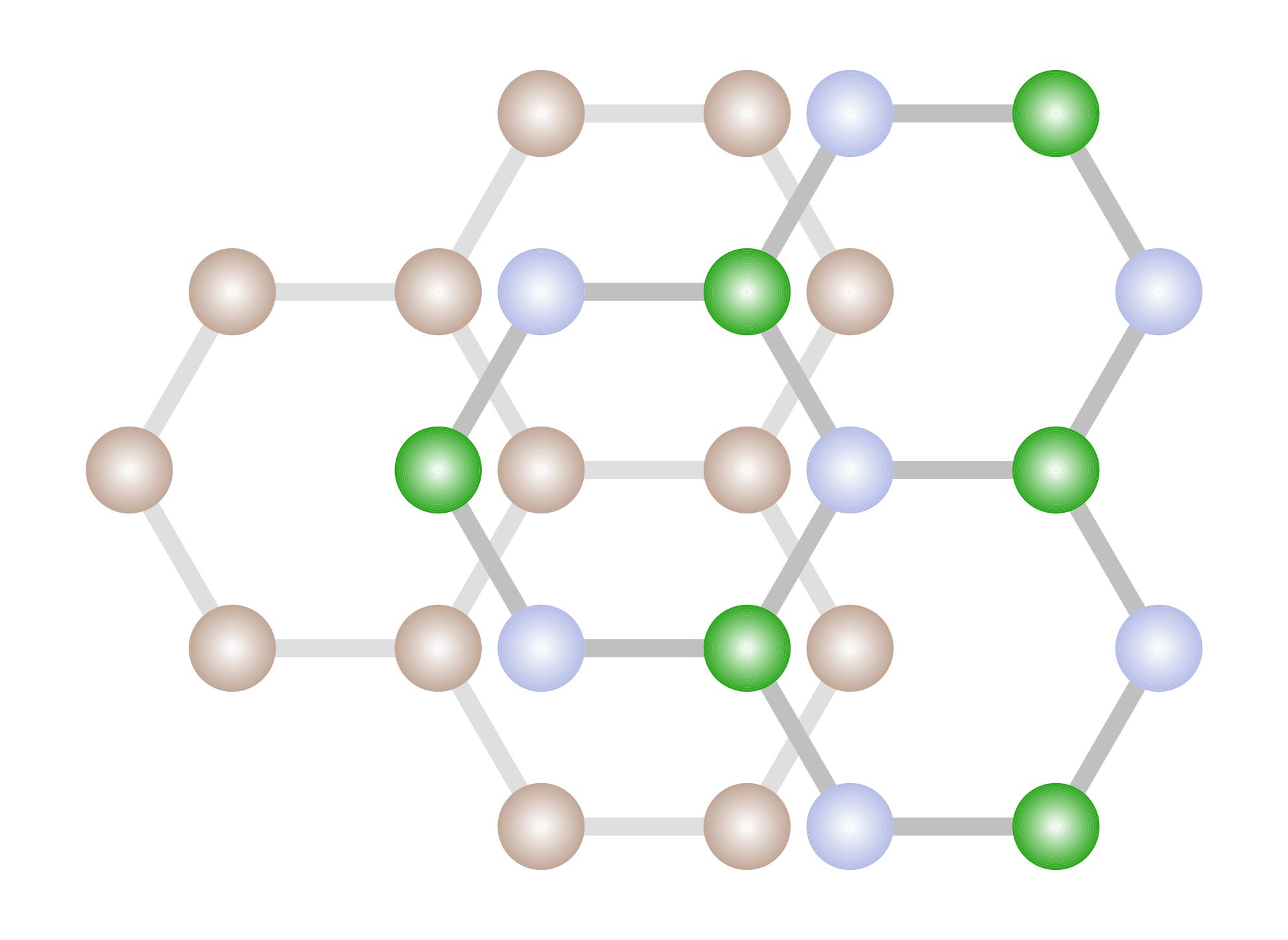}
        \end{minipage}
        \begin{minipage}[t]{0.24\textwidth}
            \centering
            BA $(s = 1/3)$ \\
            \includegraphics[height=0.9in]{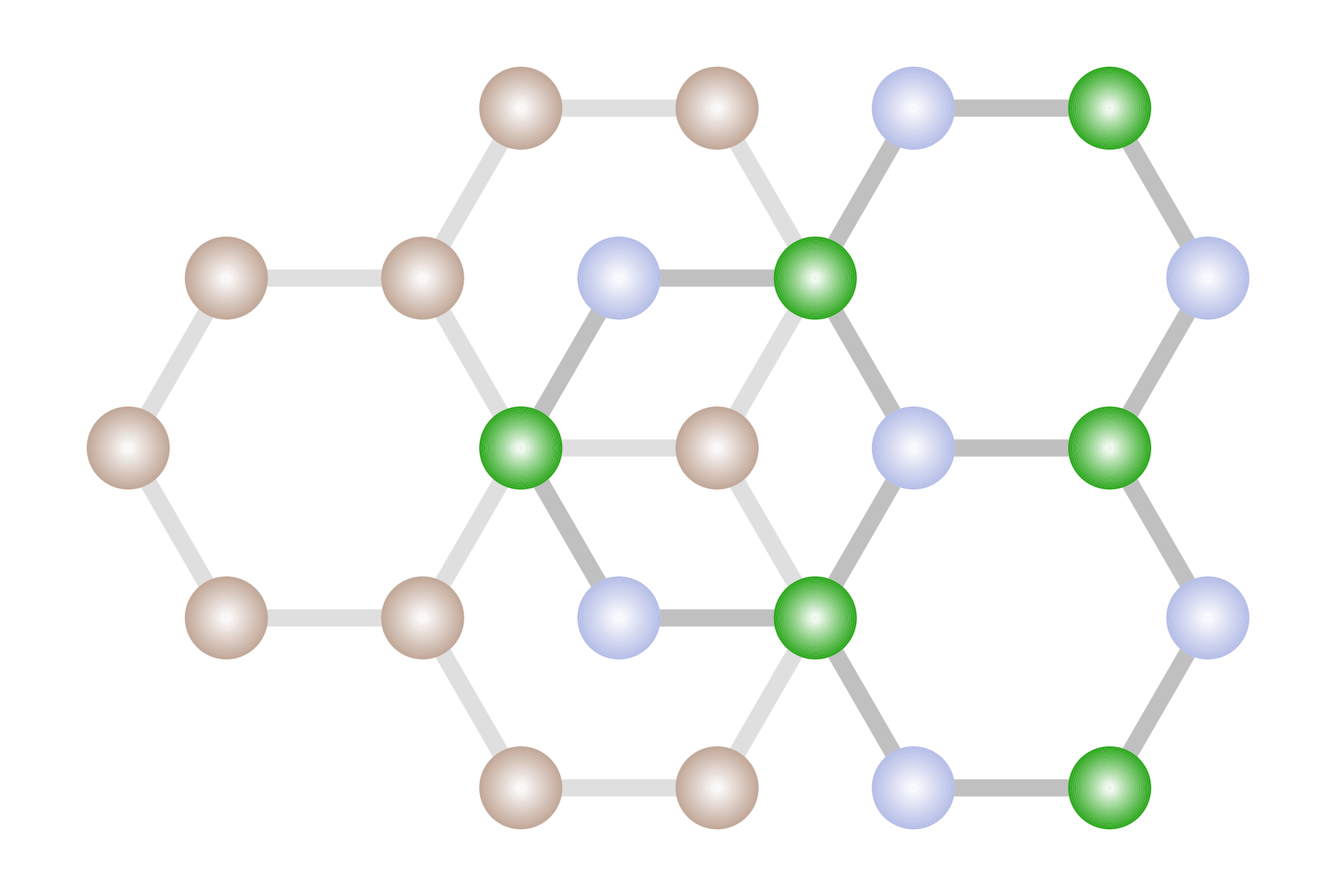}
        \end{minipage}
        \begin{minipage}[t]{0.24\textwidth}
            \centering
            AA $(s = 2/3)$ \\
            \includegraphics[height=0.9in]{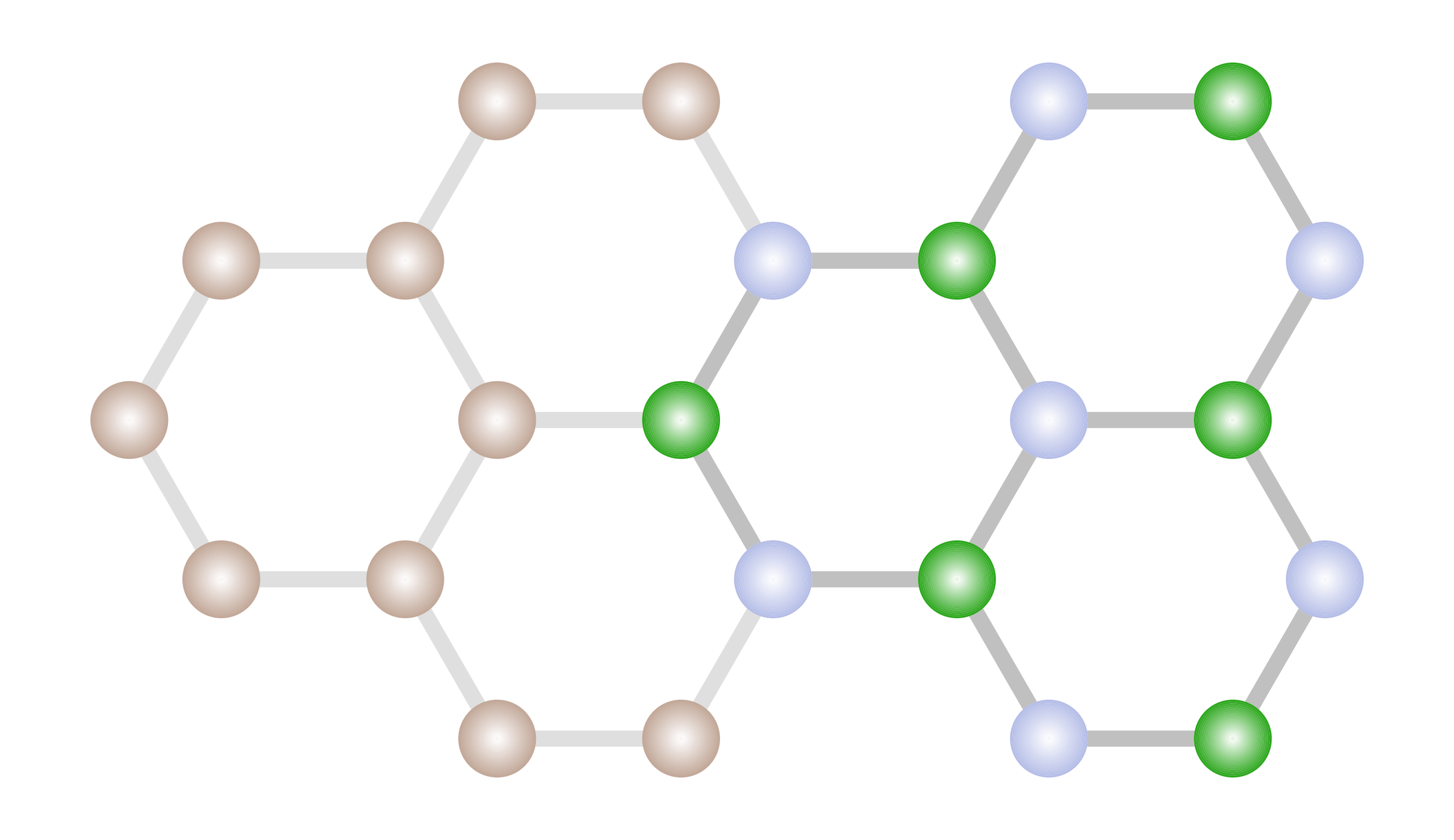}
        \end{minipage}
    \end{tabular}
    \caption{
    Stacking registries for rigid bilayer graphene and graphene-hBN bilayer.
    In both systems, the lattice constants of both layers are set to the monolayer graphene lattice constant $a_{\textrm{G}} = 2.46~\textrm{\AA}$.
    The SP, BA, Mid, and AA stacking types are defined by translating the top layer while fixing the bottom layer by $s = 1/6, 1/3, 1/2$, and $2/3$, respectively in the units of $\sqrt{3} a_{\textrm{G}}$.
    }
    \label{fig:stack}
\end{figure*}

To generate the training data set for the atomistic potential, we performed fixed-node diffusion quantum Monte Carlo as implemented in \texttt{QMCPACK}~\cite{kim2018qmcpack}.
We used three-dimensional twisted boundary conditions with a cell height of 40~\AA.
For each supercell, we evaluated the energy on a $\mathbf{k}$-grid of $4 \times 4 \times 1$.
The Slater--Jastrow-type wave function is used.
We used a two-body Jastrow factor as explained in Ref.~\cite{kim2018qmcpack}.
The Slater part of the wave function is constructed using orbitals from DFT calculation in \texttt{Quantum Espresso}~\cite{giannozzi2009quantum,giannozzi2017advanced,giannozzi2020quantum}.
The robustness of QMC is tested against different functionals.
The correlation consistent effective core potentials (ccECP)~\cite{bennett2017new} are used to remove core electrons.
The time step is set to $0.02~\mathrm{Ha}^{-1}$.
This choice was verified against the time step of $0.01~\mathrm{Ha}^{-1}$ in Ref.~\cite{krongchon2023registry} which results in errors in the energy difference less than 1~meV near the minimum.

For each interlayer distance, the energy data point is obtained from extrapolating QMC energies of $3 \times 3$, $4 \times 4$, and $5 \times 5$ (Fig.~\ref{fig:hbn_extrap}), graphene-hBN bilayer supercells to the thermodynamic limit using the equation~\cite{drummond2008finite}:
\begin{align}
E(N) = E(\infty) + cN^{-5/4}, \label{eq:hbn_qmc_extrap}
\end{align}
where $E(\infty)$ is a fitting parameter, which represents the extrapolated energy, and $N$ is the number of primitive cells in a simulation cell.
For each supercell, we performed twist averaging on the aforementioned 4$\times$ 4 k-point grid.
The statistical error bar from QMC is obtained from the bootstrapping technique.

\begin{figure*}
\includegraphics[width=7in]{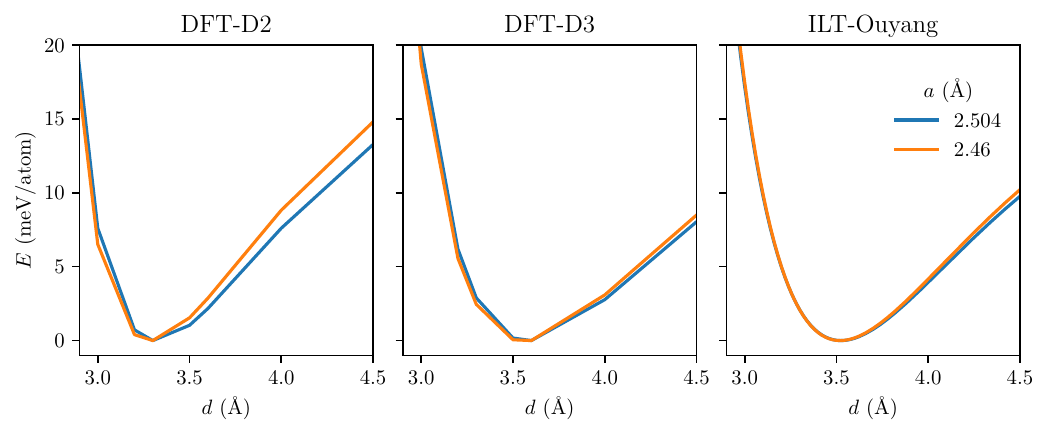}
\caption{
Interlayer interaction energy at AB stacking computed using different methods and lattice constants.
\label{fig:energy_curve_different_lattice_constants}
}
\end{figure*}

In our graphene-hBN bilayer simulation cell, hBN is compressed to the lattice constant of graphene $a_{\textrm{G}}=2.46~\textrm{\AA}$.
We tested that the effect of this compression is negligible by performing calculations at different lattice constants (Fig.~\ref{fig:inplane} and Fig.~\ref{fig:energy_curve_different_lattice_constants}).
QMC energy data points are collected from configurations AB, SP, BA, and AA stacking registries (Fig.~\ref{fig:stack}) with interlayer distances from 3 to $8~\textrm{\AA}$ (Fig.~\ref{fig:binding_curve}).

\subsection{Atomistic potential}

\subsubsection{Intralayer potential}

\begin{figure*}
    \centering
    \includegraphics{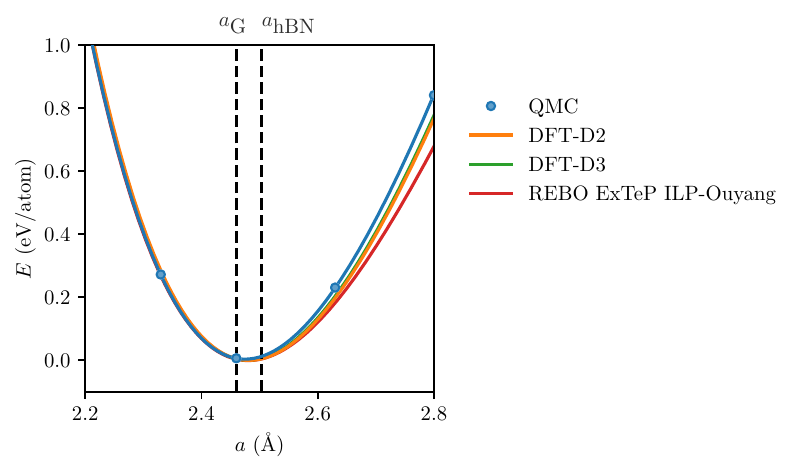}
    \caption{
    The in-plane compression and stretching of the graphene-hBN bilayer from QMC, DFT-D2, DFT-D3, and the atomistic calculation using the REBO, ExTeP, and ILP-Ouyang potentials.
    The vertical lines denote the lattice constants of monolayer graphene $a_{\textrm{G}} = 2.46~\textrm{\AA}$ and monolayer hBN $a_{\textrm{hBN}} = 2.504~\textrm{\AA}$.
    The line fit is a Morse potential.
    The statistical error bars from QMC are smaller than the markers.
    }
    \label{fig:inplane}
\end{figure*}

\begin{table}
    \caption{
    \label{tab:inplane}
    Equilibrium in-plane lattice constants $a_{\textrm{eq}}$ from QMC, DFT-D2, DFT-D3, and the atomistic calculation using the REBO, ExTeP, and ILP-Ouyang potentials.
    The minimum is calculated by fitting the Morse potential to the energy curve in Fig.~\ref{fig:inplane}.
    The QMC error bar is obtained from the bootstrapping technique.
    }
    \centering
    \begin{ruledtabular}
    \begin{tabular}{cl}
        Method & $a_{\textrm{eq}}~(\textrm{\AA})$ \\[3pt]
        \hline & \\[-2ex]
        QMC & 2.475(5) \\
        DFT-D2 & 2.481 \\
        DFT-D3 & 2.480 \\
        REBO/ExTeP/ILP-Ouyang & 2.480 \\[3pt]
    \end{tabular}
    \end{ruledtabular}
\end{table}

The atomistic potential is calculated separately between intralayer and interlayer potentials.
We use the reactive empirical bond order (REBO)~\cite{brenner2002second} potential to describe the in-plane interactions between carbon atoms.
Meanwhile, the extended Tersoff potential (ExTeP)~\cite{los2017extended} is used to describe in-plane interactions between boron and nitrogen atoms.
Since QMC provides an in-plane energy curve (Fig.~\ref{fig:inplane}) and equilibrium in-plane lattice constant $a_{\textrm{eq}}$ (Table~\ref{tab:inplane}) that are similar to those of REBO and ExTeP, we keep the same parameters as reported in their respective publications.

\subsubsection{Interlayer potential}
To describe graphene-hBN interlayer interactions, we use the ``ILP'' as implemented in \texttt{LAMMPS}~\cite{ouyang2018nanoserpents,thompson2022lammps}.
The ILP has a similar form as the Kolmogorov--Crespi potential~\cite{kolmogorov2005registry}, where the surface normal vector $\mathbf{n}_i$ at atom $i$ is calculated by averaging the cross products of vectors pointing to the three nearest neighbors.
Thus, both potentials use the same orthogonal distance:
\begin{align}
\rho_{ij}^2 &= r_{ij}^2 - (\mathbf{r}_{ij} \cdot \mathbf{n}_i)^2, \\
\rho_{ji}^2 &= r_{ij}^2 - (\mathbf{r}_{ij} \cdot \mathbf{n}_j)^2,
\end{align}
which appears in the repulsive part of the pair interaction.
The ILP and the KC potential are based on additive pair interactions:
\begin{align}
E &= \frac{1}{2} \sum_i \sum_{j \neq i} \mathrm{Tap}(r_{ij}) V_{ij}.
\end{align}
The cutoff function (referred to as the taper function) is given by
\begin{align}
\mathrm{Tap}(x_{ij}) &= 20 x_{ij}^7 - 70  x_{ij}^6 + 84  x_{ij}^5 - 35 x_{ij}^4 + 1, \\
x_{ij} &= \frac{r_{ij}}{R_{\rm{cut}}},
\end{align}
where $R_{\mathrm{cut}}$ is fixed to $16~\mathrm{\text{\AA}}$ throughout this work.

However, there are minor differences between the KC potential and the ILP in the detailed pair interaction.
The KC potential has the form
\begin{align}
V_{ij} &= e^{-\lambda(r_{ij} - z_0)} [C + f(\rho_{ij}) + f(\rho_{ji})] \\
&~~~~- A\left(\frac{r_{ij}}{z_0}\right)^{-6}. \\
f(\rho) &= e^{-(\rho/\delta)^2} \sum_{n=0}^2 C_{2n} \left(\frac{\rho}{\delta}\right)^{2n}. \label{eq:pair_kc}
\end{align}
Meanwhile, the ILP pair interaction is given by
\begin{align}
V_{ij}^{\textrm{ILP}} &= e^{-\alpha(r_{ij}/\beta - 1)}[\epsilon + f(\rho_{ij}) + f(\rho_{ji})] \nonumber \\
&~~~~- \frac{1}{1 + e^{-d\left(\frac{r_{ij}}{s_{\textrm{R}} r_{\textrm{eff}}}-1\right)}} \frac{C_6}{r_{ij}^6}. \label{eq:v_ilp} \\
f(\rho) &= C e^{-(\rho/\gamma)^2}.
\end{align}
The repulsive part now only accounts for the first term of \ref{eq:pair_kc}.
The attractive part of the ILP has its own damping function to improve the flexibility in the high slope region.
The damping function introduces an additional parameter $d$ to modify how fast the $r_{ij}^{-6}$ approaches zero as $r_{ij}$ goes to zero in the Fermi--Dirac statistics.
The multiplication of parameters $s_{\textrm{R}}$ and $r_{\textrm{eff}}$ can be considered one parameter, which determines the zero-inflection point of the damping function.

\subsection{Fitting procedure}
The damping function provided in Eq.~(\ref{eq:v_ilp}) allows us to treat $s_{\textrm{R}}$ and $r_{\textrm{eff}}$ as one parameter.
Thus, we set the value of $r_{\textrm{eff}}$ to be the starting value provided in Ref.~\cite{ouyang2018nanoserpents} and optimize only $s_{\textrm{R}}$.
Since the energy from different sources is provided up to a constant, we do not know the amount of energy shift to apply before hand.
In fact, shifting the energy is crucial in the fitting process because of the large difference in the raw energy data between QMC and the ILP.
To improve convergence, we add a learnable constant $E_{\textrm{shift}}$ to the ILP energy model, which makes nine total learnable parameters.
Thus, the fitting equation is given by
\begin{align}
E_{\textrm{QMC}} &= E_{\textrm{REBO/ExTeP}} + E_{\textrm{ILP}}(d, s; \{p_i\}) + E_{\textrm{shift}}, \label{eq:hbn_fitting}
\end{align}
where $E_{\textrm{REBO/ExTeP}}$ is the intralayer energy, $E_{\textrm{ILP}}(d, s; \{p_i\})$ is the interlayer energy, $\{p_i\}$ denotes the potential parameters, and $E_{\textrm{QMC}}$ is the raw data from QMC calculations.
The interlayer energy $E_{\textrm{ILP}}(d, s; \{p_i\})$ goes to zero as the interlayer distance $d$ goes to infinity.
The subsequent results and discussion are based on only the interlayer energy $E_{\textrm{ILP}}$, which means that $E_{\textrm{REBO/ExTeP}}$ and $E_{\textrm{shift}}$ are subtracted from both sides of Eq.~(\ref{eq:hbn_fitting}).

The fitting procedure is performed separately between graphene-hBN heterobilayer and bilayer graphene.
The trust region reflective algorithm is used to optimize the parameters.
This algorithm allows for including parameter bounds, where in this case, all the parameters (except the energy shift) have minimum of zero and no bound on their maximum.
In the calculation of SFEs, weighted fits are applied as described in Ref.~\cite{krongchon2023registry}.

\section{Results}

\begin{figure*}
    \centering
    \hspace*{-0.5in}\includegraphics{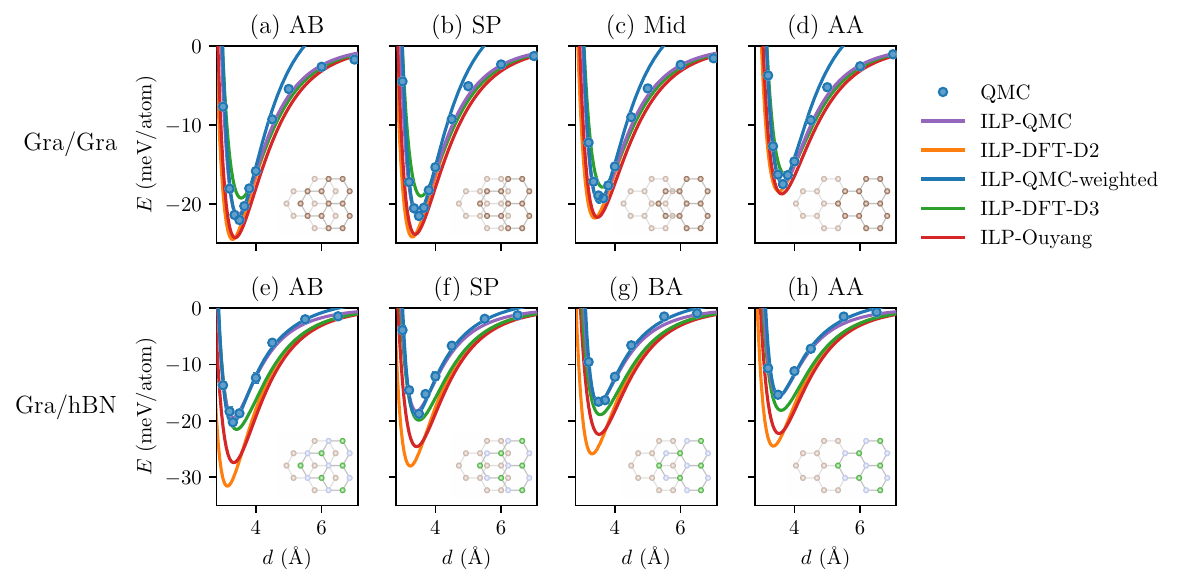}
    \caption{
    (a, b, c, d) Interlayer energy curve of bilayer graphene (Gra/Gra).
    (e, f, g, h) Interlayer energy curve of graphene-hBN heterobilayer (Gra/hBN).
    Columns represent different stacking types as defined in Fig.~\ref{fig:stack}.
    The QMC data for Gra/hBN is generated in this work.
    The QMC data for Gra/Gra is taken from Ref.~\cite{krongchon2023registry}.
    The ILP-QMC parameters are fitted in this work for both Gra/hBN and Gra/Gra.
    DFT-D2 and DFT-D3 are generated on a fine grid of small spacing and are fitted to the ILP.
    The ILP-Ouyang parameters are taken from Ref.~\cite{ouyang2018nanoserpents}.
    }
    \label{fig:binding_curve}
\end{figure*}

In Fig.~\ref{fig:binding_curve}, the top row (Fig.~\ref{fig:binding_curve}(a, b, c, d)) represents interlayer energy curve of bilayer graphene (Gra/Gra).
The bottom row (Fig.~\ref{fig:binding_curve}(e, f, g, h)) represents graphene-hBN heterobilayer (Gra/hBN).
QMC data points are plotted in blue markers.
The QMC data for Gra/Gra at AB, SP, Mid, and AA stacking types is reproduced here from Ref.~\cite{krongchon2023registry}.
The QMC data for Gra/hBN is generated in this work.
For Gra/hBN, stacking types computed are AB, SP, BA, and AA as defined in Fig.~\ref{fig:stack}.
Both sets of QMC data is fitted to the ILP potential (labeled as ILP-QMC) with $R^2 = 0.9894$ and the root-mean-square (RMS) error of $0.7~\mathrm{meV}$ in Gra/hBN.
In Gra/Gra, the fit has $R^2 = 0.9942$ and the RMS error of $0.6~\mathrm{meV}$.
The parameters are provided in Table~\ref{tab:hbn_params_qmc}.

The ILP energy curve whose parameters reported in Ref.~\cite{ouyang2018nanoserpents} (labeled as ILP-Ouyang), which is the starting point for ILP-QMC, is plotted in red.
The ILP-Ouyang parameters are also reproduced in Table~\ref{tab:hbn_params_ouyang} for reference.
The widely used DFT-D2 and DFT-D3 vdW correction methods are also included for comparison.

In Fig.~\ref{fig:binding_curve}(f), we note that the ILP-QMC fits show minor deviations from the QMC reference data, particularly in the SP stacking.
This similar behavior was previously observed in bilayer graphene~\cite{krongchon2023registry}, when the KC potential parameters were optimized, which suggests that the KC-type potential overestimates the difference between the BA stacking (the most favorable structure for a graphene-hBN bilayer) and the SP stacking.
The root-mean-square error is 1.2 meV/atom, which is approximately twice the typical QMC error bars (average 0.55 meV/atom).
To address this deviation, we performed the weighted fit based on the distance from the equilibrium point as outlined in Ref.~\cite{krongchon2023registry} in our calculations of SFE and relaxed structure. 
Therefore, we expect the impact of the fit error on applications that focus on the energetic behavior near the minimum to be negligible..

Equilibrium interlayer distance $d_{\textrm{eq}}$ and binding energy (BE) for graphene-hBN heterobilayer (Gra/hBN) and bilayer graphene (Gra/Gra) from DFT-D2, DFT-D3, ILP-Ouyang, and ILP-QMC are reported in Table~\ref{tab:hbn_be}.
In graphene-hBN heterobilayer, the BE from ILP-Ouyang deviates from QMC by roughly 6~meV across all stacking types (Fig.~\ref{fig:binding_curve}(e, f, g, h)).
This deviation is large when compared to the good agreement between ILP-Ouyang and ILP-QMC in bilayer graphene, where the deviation is only within 2~meV (Fig.~\ref{fig:binding_curve}(a, b, c, d)).
Since ILP-Ouyang was trained on DFT-MBD, this result contradicts the good performance of DFT-MBD-fitted potential (labeled as KC-Ouyang)~\cite{ouyang2018nanoserpents} in predicting the stacking-fault energy and corrugation of TBG as shown in Ref.~\cite{krongchon2023registry}.
In graphene-hBN heterobilayer, ILP-Ouyang overestimates the stacking-fault energy by predicting roughly 5~meV difference between the AB and BA stacking types, compared to 2~meV in QMC.
This prediction results in the relaxed structure incorrectly favoring the AB stacking.

In bilayer graphene (Fig.~\ref{fig:binding_curve}(a, b, c, d)), DFT-D2 and DFT-D3 show a clear trend across different stacking registries.
DFT-D2 overestimates the BE and the stacking-fault energy as observed before in QMC~\cite{mostaani2015quantum}.
In contrast, DFT-D3 underestimates the BE and the stacking-fault energy.
This underestimation of the stacking-fault energy in DFT-D3 contributes to its poor prediction of the structural corrugation in TBG.
However, in graphene-hBN heterobilayer (Fig.~\ref{fig:binding_curve}(e, f, g, h)), this trend is not observed.
In fact, despite incorrectly predicting the atomic structure in TBG, DFT-D3 is the closest match to the QMC data in Gra/hBN.
According to Table~\ref{tab:hbn_be}, the deviation in BE between DFT-D3 and ILP-QMC is roughly 2\%, whereas in the case of DFT-D2, the deviation is 11\%.

\begin{figure*}
    \centering
    \includegraphics{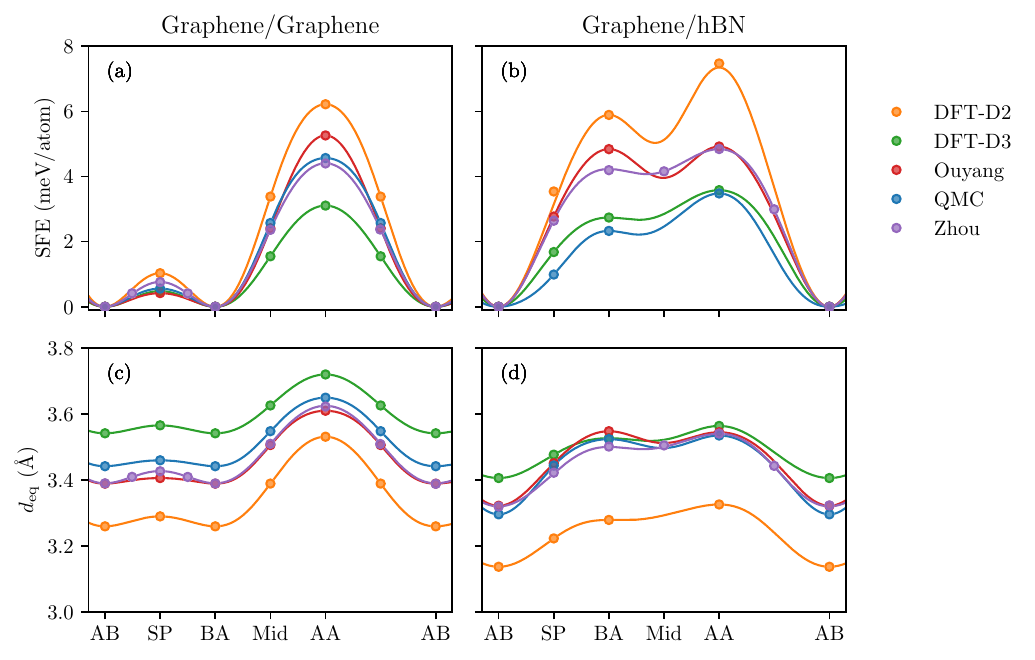}
    \caption{
    Stacking-fault energy for Graphene/Graphene and Graphene/hBN.
    ``Ouyang'' is calculated from ILP parameters provided in Ref.~\cite{ouyang2018nanoserpents}, while ``Zhou'' is taken from the random-phase approximation (RPA) results Ref.~\cite{zhou2015van}.
    Both SFE and $d_{\textrm{eq}}$ for QMC are calculated from the weighted fit in Fig.~\ref{fig:binding_curve}.
    }
    \label{fig:hbn_sfe}
\end{figure*}

Fig.~\ref{fig:hbn_sfe}(a, b) shows the SFE with respect to the minimum energy for each method.
DFT-D3 accurately predicts SFE in Graphene/hBN.
Although, Ouyang has the smallest deviation from QMC in Graphene/Graphene, it has a large deviation from QMC in Graphene/hBN.
We observe that none of the commonly used DFT-based methods being studied can accurately predict the stacking-fault energy in both Graphene/Graphene and Graphene/hBN.
The equilibrium interlayer distances $d_\text{eq}$  are similar among QMC, Ouyang, and Zhou for both systems.
In both systems, $d_\text{eq}$ is the smallest in DFT-D2, consistent with the large SFE.
On the other hand, DFT-D3 results in the highest $d_\text{eq}$ in Graphene/Graphene but generally agrees with the consensus QMC, Ouyang, and Zhou in Graphene/hBN.

\begin{table*}
    \caption{
    \label{tab:hbn_params_ouyang}
    ILP-Ouyang parameters from Ref.~\cite{ouyang2018nanoserpents}.
    }
    \centering
    \begin{tabular*}{\textwidth}{@{\extracolsep{\fill}}lrrrrrrrrr}
        \hline\hline & \\[-2ex]
        Pair & $\beta~(\textrm{\AA})$ & $\alpha$ & $\gamma~(\textrm{\AA})$ & $\epsilon~(\textrm{meV})$ & $C~(\textrm{meV})$ & $d$ & $s_{{\textrm{{R}}}}$ & $r_{{\textrm{{reff}}}}~(\textrm{\AA})$ & $C_6~(\textrm{meV} \cdot \textrm{\AA}^6)$ \\[3pt]
        \hline & \\[-2ex]
        CC & 3.2058 & 7.5111 & 1.2353 & 1.53E-05 & 37.5304 & 15.4999 & 0.7954 & 3.6814 & 25714.5 \\
        CB & 3.3037 & 10.5441 & 2.9267 & 16.7200 & 0.3572 & 15.3053 & 0.7002 & 3.0973 & 30162.9 \\
        CN & 3.2536 & 8.8259 & 1.0595 & 18.3447 & 21.9136 & 15.0000 & 0.7235 & 3.0131 & 19063.1  \\[3pt]
        \hline\hline
    \end{tabular*}
\end{table*}

\begin{table*}
    \caption{
    \label{tab:hbn_params_qmc}
    ILP parameters fitted to QMC.
    }
    \centering
    \begin{tabular*}{\textwidth}{@{\extracolsep{\fill}}lrrrrrrrrr}
        \hline\hline & \\[-2ex]
        Pair & $\beta~(\textrm{\AA})$ & $\alpha$ & $\gamma~(\textrm{\AA})$ & $\epsilon~(\textrm{meV})$ & $C~(\textrm{meV})$ & $d$ & $s_{{\textrm{{R}}}}$ & $r_{{\textrm{{reff}}}}~(\textrm{\AA})$ & $C_6~(\textrm{meV} \cdot \textrm{\AA}^6)$ \\[3pt]
        \hline & \\[-2ex]
        CC & 3.2039 & 9.4261 & 1.2535 & 0.000000 & 44.0400 & 15.4582 & 0.4183 & 3.6814 & 25936.4 \\
        CB & 2.5875 & 1.8527 & 0.5748 & 7.749360 & 5.7451 & 15.3073 & 0.5928 & 3.0973 & 30152.2 \\
        CN & 3.3049 & 11.7508 & 1.5027 & 16.98332 & 22.7457 & 15.0001 & 0.6707 & 3.0131 & 19056.3 \\[3pt]
        \hline\hline
    \end{tabular*}
\end{table*}

\section{Conclusion}
In this work, we investigated the interaction between hBN and graphene using QMC, DFT-D2, DFT-D3, and DFT-MBD-fitted ILP potential~\cite{ouyang2018nanoserpents}.
Our goal is to provide a detailed understanding of the binding energies, interlayer distances, and the influence of different stacking configurations on the structural properties of graphene-hBN heterobilayers.
By referencing accurate QMC data, our results demonstrate significant variations in binding energies and equilibrium interlayer distances across different computational methods.
DFT-D2 was found to consistently overestimate both the binding energy and the stacking-fault energy.
Notably, the DFT-D3 method, despite its inaccuracies in atomic structure prediction for TBG, showed a good agreement to QMC data for graphene-hBN systems, having a deviation in binding energy of approximately 2\%, compared to an 11\% deviation in the case of DFT-D2.
The inconsistency across different approximations highlights the importance of accurate modeling of vdW interactions to capture the correct structural and electronic properties in graphene-hBN heterobilayers.
We provide the QMC-fitted interlayer potential (ILP-QMC) as a robust tool for predicting properties such as equilibrium structures, phonons, and dynamics, in TBG on hBN or hBN-encapsulated TBG.

As we prepared this manuscript, similar result, Ref~\cite{szyniszewskiAdhesionReconstructionGraphene2025}, was published.
It appears that our QMC results are in agreement with theirs; however, we computed more points in the bonding region and fit to a more flexible potential with more realistic parameters; as the authors note, their Lennard-Jones parameters are the wrong sign. It would be interesting to study how the resulting models differ in their predictions.

\section{Acknowledgments}

This project is supoprted by U.S. Department of Energy, Office of Science, Office of Basic Energy Sciences, Computational Materials Sciences Award No. DE-SC0020177.
An award for computer time was provided by the U.S. Department of Energy’s (DOE) Innovative and Novel Computational Impact on Theory and Experiment (INCITE) Program. This research used resources from the Argonne Leadership Computing Facility, a U.S. DOE Office of Science user facility at Argonne National Laboratory, which is supported by the Office of Science of the U.S. DOE under Contract No. DE-AC02-06CH11357.
This research used resources of the National Energy Research Scientific Computing Center (NERSC), a Department of Energy User Facility (project qmc-hamm).
This work used resources of the Oak Ridge Leadership Computing Facility at the Oak Ridge National Laboratory, which is supported by the Office of Science of the U.S. Department of Energy under Contract No. DE-AC0500OR22725.

\section{Appendix}

\begin{table*}
    \caption{
        \label{tab:hbn_be}
        Equilibrium interlayer distance $d_{\textrm{eq}}$ and binding energy (BE) for graphene-hBN heterobilayer (Gra/hBN) and bilayer graphene (Gra/Gra) from DFT-D2, DFT-D3, ILP-Ouyang~\cite{ouyang2018nanoserpents}, and ILP-QMC (This work).
        For Gra/hBN, stacking types computed are AB, SP, BA, and AA.
        For Gra/Gra, stacking types computed are AB, SP, Mid, and AA.
        The equilibrium interlayer distance is calculated from the weighted fit, while the binding energy is calculated from the unweighted fit.
    }
    \centering
    \begin{tabular*}{\textwidth}{@{\extracolsep{\fill}}lllll}
        \hline\hline & \\[-2ex]
        System & Stacking & Method & $d_{\textrm{eq}}~(\textrm{\AA})$ & BE~(meV) \\[3pt]
        \hline & \\[-2ex]
        Gra/Gra & AB & ILP-Ouyang & 3.389 & 24.15 \\
        Gra/Gra & AB & ILP-QMC & 3.442 & 23.21 \\
        Gra/Gra & AB & ILP-DFT-D2 & 3.260 & 23.78 \\
        Gra/Gra & AB & ILP-DFT-D3 & 3.542 & 19.40 \\

        Gra/Gra & SP & ILP-Ouyang & 3.406 & 23.73 \\
        Gra/Gra & SP & ILP-QMC & 3.460 & 22.65 \\
        Gra/Gra & SP & ILP-DFT-D2 & 3.290 & 22.74 \\
        Gra/Gra & SP & ILP-DFT-D3 & 3.566 & 18.92 \\

        Gra/Gra & Mid & ILP-Ouyang & 3.506 & 21.74 \\
        Gra/Gra & Mid & ILP-QMC & 3.548 & 20.64 \\
        Gra/Gra & Mid & ILP-DFT-D2 & 3.389 & 20.39 \\
        Gra/Gra & Mid & ILP-DFT-D3 & 3.626 & 17.85 \\

        Gra/Gra & AA & ILP-Ouyang & 3.610 & 18.89 \\
        Gra/Gra & AA & ILP-QMC & 3.650 & 18.65 \\
        Gra/Gra & AA & ILP-DFT-D2 & 3.531 & 17.56 \\
        Gra/Gra & AA & ILP-DFT-D3 & 3.720 & 16.30 \\

        Gra/hBN & AB & ILP-DFT-D2 & 3.137 & 31.83 \\
        Gra/hBN & AB & ILP-DFT-D3 & 3.406 & 21.72 \\
        Gra/hBN & AB & ILP-QMC & 3.296 & 19.13 \\
        Gra/hBN & AB & ILP-Ouyang & 3.322 & 27.28 \\

        Gra/hBN & SP & ILP-DFT-D2 & 3.223 & 28.29 \\
        Gra/hBN & SP & ILP-DFT-D3 & 3.477 & 20.04 \\
        Gra/hBN & SP & ILP-QMC & 3.444 & 18.14 \\
        Gra/hBN & SP & ILP-Ouyang & 3.451 & 24.51 \\

        Gra/hBN & BA & ILP-DFT-D2 & 3.279 & 25.94 \\
        Gra/hBN & BA & ILP-DFT-D3 & 3.527 & 18.98 \\
        Gra/hBN & BA & ILP-QMC & 3.524 & 16.80 \\
        Gra/hBN & BA & ILP-Ouyang & 3.548 & 22.44 \\

        Gra/hBN & AA & ILP-DFT-D2 & 3.326 & 24.36 \\
        Gra/hBN & AA & ILP-DFT-D3 & 3.564 & 18.14 \\
        Gra/hBN & AA & ILP-QMC & 3.535 & 15.65 \\
        Gra/hBN & AA & ILP-Ouyang & 3.545 & 22.36 \\[3pt]
        \hline\hline
    \end{tabular*}
\end{table*}

\bibliography{paper}

\begin{thebibliography}{76}%
\makeatletter
\providecommand \@ifxundefined [1]{%
 \@ifx{#1\undefined}
}%
\providecommand \@ifnum [1]{%
 \ifnum #1\expandafter \@firstoftwo
 \else \expandafter \@secondoftwo
 \fi
}%
\providecommand \@ifx [1]{%
 \ifx #1\expandafter \@firstoftwo
 \else \expandafter \@secondoftwo
 \fi
}%
\providecommand \natexlab [1]{#1}%
\providecommand \enquote  [1]{``#1''}%
\providecommand \bibnamefont  [1]{#1}%
\providecommand \bibfnamefont [1]{#1}%
\providecommand \citenamefont [1]{#1}%
\providecommand \href@noop [0]{\@secondoftwo}%
\providecommand \href [0]{\begingroup \@sanitize@url \@href}%
\providecommand \@href[1]{\@@startlink{#1}\@@href}%
\providecommand \@@href[1]{\endgroup#1\@@endlink}%
\providecommand \@sanitize@url [0]{\catcode `\\12\catcode `\$12\catcode
  `\&12\catcode `\#12\catcode `\^12\catcode `\_12\catcode `\%12\relax}%
\providecommand \@@startlink[1]{}%
\providecommand \@@endlink[0]{}%
\providecommand \url  [0]{\begingroup\@sanitize@url \@url }%
\providecommand \@url [1]{\endgroup\@href {#1}{\urlprefix }}%
\providecommand \urlprefix  [0]{URL }%
\providecommand \Eprint [0]{\href }%
\providecommand \doibase [0]{https://doi.org/}%
\providecommand \selectlanguage [0]{\@gobble}%
\providecommand \bibinfo  [0]{\@secondoftwo}%
\providecommand \bibfield  [0]{\@secondoftwo}%
\providecommand \translation [1]{[#1]}%
\providecommand \BibitemOpen [0]{}%
\providecommand \bibitemStop [0]{}%
\providecommand \bibitemNoStop [0]{.\EOS\space}%
\providecommand \EOS [0]{\spacefactor3000\relax}%
\providecommand \BibitemShut  [1]{\csname bibitem#1\endcsname}%
\let\auto@bib@innerbib\@empty
\bibitem [{\citenamefont {{Bao}}\ \emph {et~al.}(2009)\citenamefont {{Bao}},
  \citenamefont {{Miao}}, \citenamefont {{Chen}}, \citenamefont {{Zhang}},
  \citenamefont {{Jang}}, \citenamefont {{Dames}},\ and\ \citenamefont
  {{Lau}}}]{bao2009controlled}%
  \BibitemOpen
  \bibfield  {author} {\bibinfo {author} {\bibfnamefont {W.}~\bibnamefont
  {{Bao}}}, \bibinfo {author} {\bibfnamefont {F.}~\bibnamefont {{Miao}}},
  \bibinfo {author} {\bibfnamefont {Z.}~\bibnamefont {{Chen}}}, \bibinfo
  {author} {\bibfnamefont {H.}~\bibnamefont {{Zhang}}}, \bibinfo {author}
  {\bibfnamefont {W.}~\bibnamefont {{Jang}}}, \bibinfo {author} {\bibfnamefont
  {C.}~\bibnamefont {{Dames}}},\ and\ \bibinfo {author} {\bibfnamefont {C.~N.}\
  \bibnamefont {{Lau}}},\ }\bibfield  {title} {\bibinfo {title} {{Controlled}
  ripple texturing of suspended graphene and ultrathin graphite membranes},\
  }\href {https://www.nature.com/articles/nnano.2009.191} {\bibfield  {journal}
  {\bibinfo  {journal} {{Nature} nanotechnology}\ }\textbf {\bibinfo {volume}
  {4}},\ \bibinfo {pages} {562} (\bibinfo {year} {2009})}\BibitemShut {NoStop}%
\bibitem [{\citenamefont {{Morgenstern}}(2011)}]{morgenstern2011scanning}%
  \BibitemOpen
  \bibfield  {author} {\bibinfo {author} {\bibfnamefont {M.}~\bibnamefont
  {{Morgenstern}}},\ }\bibfield  {title} {\bibinfo {title} {{Scanning}
  tunneling microscopy and spectroscopy of graphene on insulating substrates},\
  }\href {https://onlinelibrary.wiley.com/doi/abs/10.1002/pssb.201147312}
  {\bibfield  {journal} {\bibinfo  {journal} {physica status solidi (b)}\
  }\textbf {\bibinfo {volume} {248}},\ \bibinfo {pages} {2423} (\bibinfo {year}
  {2011})}\BibitemShut {NoStop}%
\bibitem [{\citenamefont {{Andrei}}\ \emph {et~al.}(2012)\citenamefont
  {{Andrei}}, \citenamefont {{Li}},\ and\ \citenamefont
  {{Du}}}]{andrei2012electronic}%
  \BibitemOpen
  \bibfield  {author} {\bibinfo {author} {\bibfnamefont {E.~Y.}\ \bibnamefont
  {{Andrei}}}, \bibinfo {author} {\bibfnamefont {G.}~\bibnamefont {{Li}}},\
  and\ \bibinfo {author} {\bibfnamefont {X.}~\bibnamefont {{Du}}},\ }\bibfield
  {title} {\bibinfo {title} {{Electronic} properties of graphene: a perspective
  from scanning tunneling microscopy and magnetotransport},\ }\href
  {https://iopscience.iop.org/article/10.1088/0034-4885/75/5/056501/}
  {\bibfield  {journal} {\bibinfo  {journal} {{Reports} on {Progress} in
  {Physics}}\ }\textbf {\bibinfo {volume} {75}},\ \bibinfo {pages} {056501}
  (\bibinfo {year} {2012})}\BibitemShut {NoStop}%
\bibitem [{\citenamefont {{Xue}}\ \emph {et~al.}(2011)\citenamefont {{Xue}},
  \citenamefont {{Sanchez}-{Yamagishi}}, \citenamefont {{Bulmash}},
  \citenamefont {{Jacquod}}, \citenamefont {{Deshpande}}, \citenamefont
  {{Watanabe}}, \citenamefont {{Taniguchi}}, \citenamefont
  {{Jarillo}-{Herrero}},\ and\ \citenamefont {LeRoy}}]{xue2011scanning}%
  \BibitemOpen
  \bibfield  {author} {\bibinfo {author} {\bibfnamefont {J.}~\bibnamefont
  {{Xue}}}, \bibinfo {author} {\bibfnamefont {J.}~\bibnamefont
  {{Sanchez}-{Yamagishi}}}, \bibinfo {author} {\bibfnamefont {D.}~\bibnamefont
  {{Bulmash}}}, \bibinfo {author} {\bibfnamefont {P.}~\bibnamefont
  {{Jacquod}}}, \bibinfo {author} {\bibfnamefont {A.}~\bibnamefont
  {{Deshpande}}}, \bibinfo {author} {\bibfnamefont {K.}~\bibnamefont
  {{Watanabe}}}, \bibinfo {author} {\bibfnamefont {T.}~\bibnamefont
  {{Taniguchi}}}, \bibinfo {author} {\bibfnamefont {P.}~\bibnamefont
  {{Jarillo}-{Herrero}}},\ and\ \bibinfo {author} {\bibfnamefont {B.~J.}\
  \bibnamefont {LeRoy}},\ }\bibfield  {title} {\bibinfo {title} {{Scanning}
  tunnelling microscopy and spectroscopy of ultra-flat graphene on hexagonal
  boron nitride},\ }\href {https://www.nature.com/articles/nmat2968} {\bibfield
   {journal} {\bibinfo  {journal} {{Nature} materials}\ }\textbf {\bibinfo
  {volume} {10}},\ \bibinfo {pages} {282} (\bibinfo {year} {2011})}\BibitemShut
  {NoStop}%
\bibitem [{\citenamefont {Novoselov}\ \emph {et~al.}(2005)\citenamefont
  {Novoselov}, \citenamefont {Jiang}, \citenamefont {Schedin}, \citenamefont
  {Booth}, \citenamefont {Khotkevich}, \citenamefont {Morozov},\ and\
  \citenamefont {Geim}}]{novoselov2005two}%
  \BibitemOpen
  \bibfield  {author} {\bibinfo {author} {\bibfnamefont {K.~S.}\ \bibnamefont
  {Novoselov}}, \bibinfo {author} {\bibfnamefont {D.}~\bibnamefont {Jiang}},
  \bibinfo {author} {\bibfnamefont {F.}~\bibnamefont {Schedin}}, \bibinfo
  {author} {\bibfnamefont {T.}~\bibnamefont {Booth}}, \bibinfo {author}
  {\bibfnamefont {V.}~\bibnamefont {Khotkevich}}, \bibinfo {author}
  {\bibfnamefont {S.}~\bibnamefont {Morozov}},\ and\ \bibinfo {author}
  {\bibfnamefont {A.~K.}\ \bibnamefont {Geim}},\ }\bibfield  {title} {\bibinfo
  {title} {Two-dimensional atomic crystals},\ }\href
  {https://www.pnas.org/doi/abs/10.1073/pnas.0502848102} {\bibfield  {journal}
  {\bibinfo  {journal} {Proceedings of the National Academy of Sciences}\
  }\textbf {\bibinfo {volume} {102}},\ \bibinfo {pages} {10451} (\bibinfo
  {year} {2005})}\BibitemShut {NoStop}%
\bibitem [{\citenamefont {Gorbachev}\ \emph {et~al.}(2011)\citenamefont
  {Gorbachev}, \citenamefont {Riaz}, \citenamefont {Nair}, \citenamefont
  {Jalil}, \citenamefont {Britnell}, \citenamefont {Belle}, \citenamefont
  {Hill}, \citenamefont {Novoselov}, \citenamefont {Watanabe}, \citenamefont
  {Taniguchi} \emph {et~al.}}]{gorbachev2010hunting}%
  \BibitemOpen
  \bibfield  {author} {\bibinfo {author} {\bibfnamefont {R.~V.}\ \bibnamefont
  {Gorbachev}}, \bibinfo {author} {\bibfnamefont {I.}~\bibnamefont {Riaz}},
  \bibinfo {author} {\bibfnamefont {R.~R.}\ \bibnamefont {Nair}}, \bibinfo
  {author} {\bibfnamefont {R.}~\bibnamefont {Jalil}}, \bibinfo {author}
  {\bibfnamefont {L.}~\bibnamefont {Britnell}}, \bibinfo {author}
  {\bibfnamefont {B.~D.}\ \bibnamefont {Belle}}, \bibinfo {author}
  {\bibfnamefont {E.~W.}\ \bibnamefont {Hill}}, \bibinfo {author}
  {\bibfnamefont {K.~S.}\ \bibnamefont {Novoselov}}, \bibinfo {author}
  {\bibfnamefont {K.}~\bibnamefont {Watanabe}}, \bibinfo {author}
  {\bibfnamefont {T.}~\bibnamefont {Taniguchi}}, \emph {et~al.},\ }\bibfield
  {title} {\bibinfo {title} {Hunting for monolayer boron nitride: optical and
  raman signatures},\ }\href
  {https://onlinelibrary.wiley.com/doi/10.1002/smll.201001628} {\bibfield
  {journal} {\bibinfo  {journal} {Small}\ } (\bibinfo {year}
  {2011})}\BibitemShut {NoStop}%
\bibitem [{\citenamefont {Zunger}\ \emph {et~al.}(1976)\citenamefont {Zunger},
  \citenamefont {Katzir},\ and\ \citenamefont {Halperin}}]{zunger1976optical}%
  \BibitemOpen
  \bibfield  {author} {\bibinfo {author} {\bibfnamefont {A.}~\bibnamefont
  {Zunger}}, \bibinfo {author} {\bibfnamefont {A.}~\bibnamefont {Katzir}},\
  and\ \bibinfo {author} {\bibfnamefont {A.}~\bibnamefont {Halperin}},\
  }\bibfield  {title} {\bibinfo {title} {Optical properties of hexagonal boron
  nitride},\ }\href
  {https://journals.aps.org/prb/abstract/10.1103/PhysRevB.13.5560} {\bibfield
  {journal} {\bibinfo  {journal} {Physical Review B}\ }\textbf {\bibinfo
  {volume} {13}},\ \bibinfo {pages} {5560} (\bibinfo {year}
  {1976})}\BibitemShut {NoStop}%
\bibitem [{\citenamefont {Watanabe}\ \emph {et~al.}(2004)\citenamefont
  {Watanabe}, \citenamefont {Taniguchi},\ and\ \citenamefont
  {Kanda}}]{watanabe2004direct}%
  \BibitemOpen
  \bibfield  {author} {\bibinfo {author} {\bibfnamefont {K.}~\bibnamefont
  {Watanabe}}, \bibinfo {author} {\bibfnamefont {T.}~\bibnamefont
  {Taniguchi}},\ and\ \bibinfo {author} {\bibfnamefont {H.}~\bibnamefont
  {Kanda}},\ }\bibfield  {title} {\bibinfo {title} {Direct-bandgap properties
  and evidence for ultraviolet lasing of hexagonal boron nitride single
  crystal},\ }\href {https://www.nature.com/articles/nmat1134} {\bibfield
  {journal} {\bibinfo  {journal} {Nature materials}\ }\textbf {\bibinfo
  {volume} {3}},\ \bibinfo {pages} {404} (\bibinfo {year} {2004})}\BibitemShut
  {NoStop}%
\bibitem [{\citenamefont {Yankowitz}\ \emph
  {et~al.}(2019{\natexlab{a}})\citenamefont {Yankowitz}, \citenamefont {Ma},
  \citenamefont {Jarillo-Herrero},\ and\ \citenamefont
  {LeRoy}}]{yankowitz2019van}%
  \BibitemOpen
  \bibfield  {author} {\bibinfo {author} {\bibfnamefont {M.}~\bibnamefont
  {Yankowitz}}, \bibinfo {author} {\bibfnamefont {Q.}~\bibnamefont {Ma}},
  \bibinfo {author} {\bibfnamefont {P.}~\bibnamefont {Jarillo-Herrero}},\ and\
  \bibinfo {author} {\bibfnamefont {B.~J.}\ \bibnamefont {LeRoy}},\ }\bibfield
  {title} {\bibinfo {title} {van der waals heterostructures combining graphene
  and hexagonal boron nitride},\ }\href
  {https://www.nature.com/articles/s42254-018-0016-0} {\bibfield  {journal}
  {\bibinfo  {journal} {Nature Reviews Physics}\ }\textbf {\bibinfo {volume}
  {1}},\ \bibinfo {pages} {112} (\bibinfo {year}
  {2019}{\natexlab{a}})}\BibitemShut {NoStop}%
\bibitem [{\citenamefont {Elias}\ \emph {et~al.}(2019)\citenamefont {Elias},
  \citenamefont {Valvin}, \citenamefont {Pelini}, \citenamefont {Summerfield},
  \citenamefont {Mellor}, \citenamefont {Cheng}, \citenamefont {Eaves},
  \citenamefont {Foxon}, \citenamefont {Beton}, \citenamefont {Novikov} \emph
  {et~al.}}]{elias2019direct}%
  \BibitemOpen
  \bibfield  {author} {\bibinfo {author} {\bibfnamefont {C.}~\bibnamefont
  {Elias}}, \bibinfo {author} {\bibfnamefont {P.}~\bibnamefont {Valvin}},
  \bibinfo {author} {\bibfnamefont {T.}~\bibnamefont {Pelini}}, \bibinfo
  {author} {\bibfnamefont {A.}~\bibnamefont {Summerfield}}, \bibinfo {author}
  {\bibfnamefont {C.}~\bibnamefont {Mellor}}, \bibinfo {author} {\bibfnamefont
  {T.}~\bibnamefont {Cheng}}, \bibinfo {author} {\bibfnamefont
  {L.}~\bibnamefont {Eaves}}, \bibinfo {author} {\bibfnamefont
  {C.}~\bibnamefont {Foxon}}, \bibinfo {author} {\bibfnamefont
  {P.}~\bibnamefont {Beton}}, \bibinfo {author} {\bibfnamefont
  {S.}~\bibnamefont {Novikov}}, \emph {et~al.},\ }\bibfield  {title} {\bibinfo
  {title} {Direct band-gap crossover in epitaxial monolayer boron nitride},\
  }\href {https://www.nature.com/articles/s41467-019-10610-5} {\bibfield
  {journal} {\bibinfo  {journal} {Nature communications}\ }\textbf {\bibinfo
  {volume} {10}},\ \bibinfo {pages} {2639} (\bibinfo {year}
  {2019})}\BibitemShut {NoStop}%
\bibitem [{\citenamefont {{Qiu}}\ \emph {et~al.}(2017)\citenamefont {{Qiu}},
  \citenamefont {da~{Jornada}},\ and\ \citenamefont
  {{Louie}}}]{qiu2017environmental}%
  \BibitemOpen
  \bibfield  {author} {\bibinfo {author} {\bibfnamefont {D.~Y.}\ \bibnamefont
  {{Qiu}}}, \bibinfo {author} {\bibfnamefont {F.~H.}\ \bibnamefont
  {da~{Jornada}}},\ and\ \bibinfo {author} {\bibfnamefont {S.~G.}\ \bibnamefont
  {{Louie}}},\ }\bibfield  {title} {\bibinfo {title} {{Environmental} screening
  effects in {2D} materials: renormalization of the bandgap, electronic
  structure, and optical spectra of few-layer black phosphorus},\ }\href
  {https://pubs.acs.org/doi/abs/10.1021/acs.nanolett.7b01365} {\bibfield
  {journal} {\bibinfo  {journal} {{Nano} letters}\ }\textbf {\bibinfo {volume}
  {17}},\ \bibinfo {pages} {4706} (\bibinfo {year} {2017})}\BibitemShut
  {NoStop}%
\bibitem [{\citenamefont {Dean}\ \emph {et~al.}(2010)\citenamefont {Dean},
  \citenamefont {Young}, \citenamefont {Meric}, \citenamefont {Lee},
  \citenamefont {Wang}, \citenamefont {Sorgenfrei}, \citenamefont {Watanabe},
  \citenamefont {Taniguchi}, \citenamefont {Kim}, \citenamefont {Shepard} \emph
  {et~al.}}]{dean2010boron}%
  \BibitemOpen
  \bibfield  {author} {\bibinfo {author} {\bibfnamefont {C.~R.}\ \bibnamefont
  {Dean}}, \bibinfo {author} {\bibfnamefont {A.~F.}\ \bibnamefont {Young}},
  \bibinfo {author} {\bibfnamefont {I.}~\bibnamefont {Meric}}, \bibinfo
  {author} {\bibfnamefont {C.}~\bibnamefont {Lee}}, \bibinfo {author}
  {\bibfnamefont {L.}~\bibnamefont {Wang}}, \bibinfo {author} {\bibfnamefont
  {S.}~\bibnamefont {Sorgenfrei}}, \bibinfo {author} {\bibfnamefont
  {K.}~\bibnamefont {Watanabe}}, \bibinfo {author} {\bibfnamefont
  {T.}~\bibnamefont {Taniguchi}}, \bibinfo {author} {\bibfnamefont
  {P.}~\bibnamefont {Kim}}, \bibinfo {author} {\bibfnamefont {K.~L.}\
  \bibnamefont {Shepard}}, \emph {et~al.},\ }\bibfield  {title} {\bibinfo
  {title} {Boron nitride substrates for high-quality graphene electronics},\
  }\href {https://www.nature.com/articles/nnano.2010.172} {\bibfield  {journal}
  {\bibinfo  {journal} {Nature nanotechnology}\ }\textbf {\bibinfo {volume}
  {5}},\ \bibinfo {pages} {722} (\bibinfo {year} {2010})}\BibitemShut {NoStop}%
\bibitem [{\citenamefont {Hunt}\ \emph {et~al.}(2013)\citenamefont {Hunt},
  \citenamefont {Sanchez-Yamagishi}, \citenamefont {Young}, \citenamefont
  {Yankowitz}, \citenamefont {LeRoy}, \citenamefont {Watanabe}, \citenamefont
  {Taniguchi}, \citenamefont {Moon}, \citenamefont {Koshino}, \citenamefont
  {Jarillo-Herrero} \emph {et~al.}}]{hunt2013massive}%
  \BibitemOpen
  \bibfield  {author} {\bibinfo {author} {\bibfnamefont {B.}~\bibnamefont
  {Hunt}}, \bibinfo {author} {\bibfnamefont {J.~D.}\ \bibnamefont
  {Sanchez-Yamagishi}}, \bibinfo {author} {\bibfnamefont {A.~F.}\ \bibnamefont
  {Young}}, \bibinfo {author} {\bibfnamefont {M.}~\bibnamefont {Yankowitz}},
  \bibinfo {author} {\bibfnamefont {B.~J.}\ \bibnamefont {LeRoy}}, \bibinfo
  {author} {\bibfnamefont {K.}~\bibnamefont {Watanabe}}, \bibinfo {author}
  {\bibfnamefont {T.}~\bibnamefont {Taniguchi}}, \bibinfo {author}
  {\bibfnamefont {P.}~\bibnamefont {Moon}}, \bibinfo {author} {\bibfnamefont
  {M.}~\bibnamefont {Koshino}}, \bibinfo {author} {\bibfnamefont
  {P.}~\bibnamefont {Jarillo-Herrero}}, \emph {et~al.},\ }\bibfield  {title}
  {\bibinfo {title} {Massive dirac fermions and hofstadter butterfly in a van
  der waals heterostructure},\ }\href {https://doi.org/10.1126/science.1237240}
  {\bibfield  {journal} {\bibinfo  {journal} {Science}\ }\textbf {\bibinfo
  {volume} {340}},\ \bibinfo {pages} {1427} (\bibinfo {year}
  {2013})}\BibitemShut {NoStop}%
\bibitem [{\citenamefont {Geim}\ and\ \citenamefont
  {Grigorieva}(2013)}]{geim2013van}%
  \BibitemOpen
  \bibfield  {author} {\bibinfo {author} {\bibfnamefont {A.~K.}\ \bibnamefont
  {Geim}}\ and\ \bibinfo {author} {\bibfnamefont {I.~V.}\ \bibnamefont
  {Grigorieva}},\ }\bibfield  {title} {\bibinfo {title} {Van der waals
  heterostructures},\ }\href {https://www.nature.com/articles/nature12385}
  {\bibfield  {journal} {\bibinfo  {journal} {Nature}\ }\textbf {\bibinfo
  {volume} {499}},\ \bibinfo {pages} {419} (\bibinfo {year}
  {2013})}\BibitemShut {NoStop}%
\bibitem [{\citenamefont {Aggoune}\ \emph {et~al.}(2017)\citenamefont
  {Aggoune}, \citenamefont {Cocchi}, \citenamefont {Nabok}, \citenamefont
  {Rezouali}, \citenamefont {Akli~Belkhir},\ and\ \citenamefont
  {Draxl}}]{aggoune2017enhanced}%
  \BibitemOpen
  \bibfield  {author} {\bibinfo {author} {\bibfnamefont {W.}~\bibnamefont
  {Aggoune}}, \bibinfo {author} {\bibfnamefont {C.}~\bibnamefont {Cocchi}},
  \bibinfo {author} {\bibfnamefont {D.}~\bibnamefont {Nabok}}, \bibinfo
  {author} {\bibfnamefont {K.}~\bibnamefont {Rezouali}}, \bibinfo {author}
  {\bibfnamefont {M.}~\bibnamefont {Akli~Belkhir}},\ and\ \bibinfo {author}
  {\bibfnamefont {C.}~\bibnamefont {Draxl}},\ }\bibfield  {title} {\bibinfo
  {title} {Enhanced light--matter interaction in graphene/h-bn van der waals
  heterostructures},\ }\href
  {https://pubs.acs.org/doi/10.1021/acs.jpclett.7b00357} {\bibfield  {journal}
  {\bibinfo  {journal} {The Journal of Physical Chemistry Letters}\ }\textbf
  {\bibinfo {volume} {8}},\ \bibinfo {pages} {1464} (\bibinfo {year}
  {2017})}\BibitemShut {NoStop}%
\bibitem [{\citenamefont {Withers}\ \emph {et~al.}(2015)\citenamefont
  {Withers}, \citenamefont {Del Pozo-Zamudio}, \citenamefont {Mishchenko},
  \citenamefont {Rooney}, \citenamefont {Gholinia}, \citenamefont {Watanabe},
  \citenamefont {Taniguchi}, \citenamefont {Haigh}, \citenamefont {Geim},
  \citenamefont {Tartakovskii} \emph {et~al.}}]{withers2015light}%
  \BibitemOpen
  \bibfield  {author} {\bibinfo {author} {\bibfnamefont {F.}~\bibnamefont
  {Withers}}, \bibinfo {author} {\bibfnamefont {O.}~\bibnamefont {Del
  Pozo-Zamudio}}, \bibinfo {author} {\bibfnamefont {A.}~\bibnamefont
  {Mishchenko}}, \bibinfo {author} {\bibfnamefont {A.~P.}\ \bibnamefont
  {Rooney}}, \bibinfo {author} {\bibfnamefont {A.}~\bibnamefont {Gholinia}},
  \bibinfo {author} {\bibfnamefont {K.}~\bibnamefont {Watanabe}}, \bibinfo
  {author} {\bibfnamefont {T.}~\bibnamefont {Taniguchi}}, \bibinfo {author}
  {\bibfnamefont {S.~J.}\ \bibnamefont {Haigh}}, \bibinfo {author}
  {\bibfnamefont {A.}~\bibnamefont {Geim}}, \bibinfo {author} {\bibfnamefont
  {A.}~\bibnamefont {Tartakovskii}}, \emph {et~al.},\ }\bibfield  {title}
  {\bibinfo {title} {Light-emitting diodes by band-structure engineering in van
  der waals heterostructures},\ }\href
  {https://www.nature.com/articles/nmat4205} {\bibfield  {journal} {\bibinfo
  {journal} {Nature materials}\ }\textbf {\bibinfo {volume} {14}},\ \bibinfo
  {pages} {301} (\bibinfo {year} {2015})}\BibitemShut {NoStop}%
\bibitem [{\citenamefont {Kamalakar}\ \emph {et~al.}(2014)\citenamefont
  {Kamalakar}, \citenamefont {Dankert}, \citenamefont {Bergsten}, \citenamefont
  {Ive},\ and\ \citenamefont {Dash}}]{kamalakar2014spintronics}%
  \BibitemOpen
  \bibfield  {author} {\bibinfo {author} {\bibfnamefont {M.~V.}\ \bibnamefont
  {Kamalakar}}, \bibinfo {author} {\bibfnamefont {A.}~\bibnamefont {Dankert}},
  \bibinfo {author} {\bibfnamefont {J.}~\bibnamefont {Bergsten}}, \bibinfo
  {author} {\bibfnamefont {T.}~\bibnamefont {Ive}},\ and\ \bibinfo {author}
  {\bibfnamefont {S.~P.}\ \bibnamefont {Dash}},\ }\bibfield  {title} {\bibinfo
  {title} {Spintronics with graphene-hexagonal boron nitride van der waals
  heterostructures},\ }\href {https://doi.org/10.1063/1.4902814} {\bibfield
  {journal} {\bibinfo  {journal} {Applied Physics Letters}\ }\textbf {\bibinfo
  {volume} {105}} (\bibinfo {year} {2014})}\BibitemShut {NoStop}%
\bibitem [{\citenamefont {He}\ \emph {et~al.}(2014)\citenamefont {He},
  \citenamefont {Tsutsui}, \citenamefont {Ryuzaki}, \citenamefont {Yokota},
  \citenamefont {Taniguchi},\ and\ \citenamefont {Kawai}}]{he2014graphene}%
  \BibitemOpen
  \bibfield  {author} {\bibinfo {author} {\bibfnamefont {Y.}~\bibnamefont
  {He}}, \bibinfo {author} {\bibfnamefont {M.}~\bibnamefont {Tsutsui}},
  \bibinfo {author} {\bibfnamefont {S.}~\bibnamefont {Ryuzaki}}, \bibinfo
  {author} {\bibfnamefont {K.}~\bibnamefont {Yokota}}, \bibinfo {author}
  {\bibfnamefont {M.}~\bibnamefont {Taniguchi}},\ and\ \bibinfo {author}
  {\bibfnamefont {T.}~\bibnamefont {Kawai}},\ }\bibfield  {title} {\bibinfo
  {title} {Graphene/hexagonal boron nitride/graphene nanopore for electrical
  detection of single molecules},\ }\href
  {https://www.nature.com/articles/am201429} {\bibfield  {journal} {\bibinfo
  {journal} {NPG Asia materials}\ }\textbf {\bibinfo {volume} {6}},\ \bibinfo
  {pages} {e104} (\bibinfo {year} {2014})}\BibitemShut {NoStop}%
\bibitem [{\citenamefont {Shukla}\ \emph {et~al.}(2017)\citenamefont {Shukla},
  \citenamefont {Jena}, \citenamefont {Grigoriev},\ and\ \citenamefont
  {Ahuja}}]{shukla2017prospects}%
  \BibitemOpen
  \bibfield  {author} {\bibinfo {author} {\bibfnamefont {V.}~\bibnamefont
  {Shukla}}, \bibinfo {author} {\bibfnamefont {N.~K.}\ \bibnamefont {Jena}},
  \bibinfo {author} {\bibfnamefont {A.}~\bibnamefont {Grigoriev}},\ and\
  \bibinfo {author} {\bibfnamefont {R.}~\bibnamefont {Ahuja}},\ }\bibfield
  {title} {\bibinfo {title} {Prospects of graphene--hbn heterostructure nanogap
  for dna sequencing},\ }\href
  {https://pubs.acs.org/doi/10.1021/acsami.7b06827} {\bibfield  {journal}
  {\bibinfo  {journal} {ACS applied materials \& interfaces}\ }\textbf
  {\bibinfo {volume} {9}},\ \bibinfo {pages} {39945} (\bibinfo {year}
  {2017})}\BibitemShut {NoStop}%
\bibitem [{\citenamefont {{\v{S}}i{\v{s}}kins}\ \emph
  {et~al.}(2019)\citenamefont {{\v{S}}i{\v{s}}kins}, \citenamefont {Mullan},
  \citenamefont {Son}, \citenamefont {Yin}, \citenamefont {Watanabe},
  \citenamefont {Taniguchi}, \citenamefont {Ghazaryan}, \citenamefont
  {Novoselov},\ and\ \citenamefont {Mishchenko}}]{vsivskins2019high}%
  \BibitemOpen
  \bibfield  {author} {\bibinfo {author} {\bibfnamefont {M.}~\bibnamefont
  {{\v{S}}i{\v{s}}kins}}, \bibinfo {author} {\bibfnamefont {C.}~\bibnamefont
  {Mullan}}, \bibinfo {author} {\bibfnamefont {S.-K.}\ \bibnamefont {Son}},
  \bibinfo {author} {\bibfnamefont {J.}~\bibnamefont {Yin}}, \bibinfo {author}
  {\bibfnamefont {K.}~\bibnamefont {Watanabe}}, \bibinfo {author}
  {\bibfnamefont {T.}~\bibnamefont {Taniguchi}}, \bibinfo {author}
  {\bibfnamefont {D.}~\bibnamefont {Ghazaryan}}, \bibinfo {author}
  {\bibfnamefont {K.~S.}\ \bibnamefont {Novoselov}},\ and\ \bibinfo {author}
  {\bibfnamefont {A.}~\bibnamefont {Mishchenko}},\ }\bibfield  {title}
  {\bibinfo {title} {High-temperature electronic devices enabled by
  hbn-encapsulated graphene},\ }\href
  {https://pubs.aip.org/aip/apl/article-abstract/114/12/123104/311431}
  {\bibfield  {journal} {\bibinfo  {journal} {Applied Physics Letters}\
  }\textbf {\bibinfo {volume} {114}} (\bibinfo {year} {2019})}\BibitemShut
  {NoStop}%
\bibitem [{\citenamefont {Song}\ \emph {et~al.}(2013)\citenamefont {Song},
  \citenamefont {Shytov},\ and\ \citenamefont {Levitov}}]{song2013electron}%
  \BibitemOpen
  \bibfield  {author} {\bibinfo {author} {\bibfnamefont {J.~C.}\ \bibnamefont
  {Song}}, \bibinfo {author} {\bibfnamefont {A.~V.}\ \bibnamefont {Shytov}},\
  and\ \bibinfo {author} {\bibfnamefont {L.~S.}\ \bibnamefont {Levitov}},\
  }\bibfield  {title} {\bibinfo {title} {Electron interactions and gap opening
  in graphene superlattices},\ }\href
  {https://journals.aps.org/prl/abstract/10.1103/PhysRevLett.111.266801}
  {\bibfield  {journal} {\bibinfo  {journal} {Physical review letters}\
  }\textbf {\bibinfo {volume} {111}},\ \bibinfo {pages} {266801} (\bibinfo
  {year} {2013})}\BibitemShut {NoStop}%
\bibitem [{\citenamefont {Yankowitz}\ \emph {et~al.}(2012)\citenamefont
  {Yankowitz}, \citenamefont {Xue}, \citenamefont {Cormode}, \citenamefont
  {Sanchez-Yamagishi}, \citenamefont {Watanabe}, \citenamefont {Taniguchi},
  \citenamefont {Jarillo-Herrero}, \citenamefont {Jacquod},\ and\ \citenamefont
  {LeRoy}}]{yankowitz2012emergence}%
  \BibitemOpen
  \bibfield  {author} {\bibinfo {author} {\bibfnamefont {M.}~\bibnamefont
  {Yankowitz}}, \bibinfo {author} {\bibfnamefont {J.}~\bibnamefont {Xue}},
  \bibinfo {author} {\bibfnamefont {D.}~\bibnamefont {Cormode}}, \bibinfo
  {author} {\bibfnamefont {J.~D.}\ \bibnamefont {Sanchez-Yamagishi}}, \bibinfo
  {author} {\bibfnamefont {K.}~\bibnamefont {Watanabe}}, \bibinfo {author}
  {\bibfnamefont {T.}~\bibnamefont {Taniguchi}}, \bibinfo {author}
  {\bibfnamefont {P.}~\bibnamefont {Jarillo-Herrero}}, \bibinfo {author}
  {\bibfnamefont {P.}~\bibnamefont {Jacquod}},\ and\ \bibinfo {author}
  {\bibfnamefont {B.~J.}\ \bibnamefont {LeRoy}},\ }\bibfield  {title} {\bibinfo
  {title} {Emergence of superlattice dirac points in graphene on hexagonal
  boron nitride},\ }\href {https://www.nature.com/articles/nphys2272}
  {\bibfield  {journal} {\bibinfo  {journal} {Nature physics}\ }\textbf
  {\bibinfo {volume} {8}},\ \bibinfo {pages} {382} (\bibinfo {year}
  {2012})}\BibitemShut {NoStop}%
\bibitem [{\citenamefont {Ponomarenko}\ \emph {et~al.}(2013)\citenamefont
  {Ponomarenko}, \citenamefont {Gorbachev}, \citenamefont {Yu}, \citenamefont
  {Elias}, \citenamefont {Jalil}, \citenamefont {Patel}, \citenamefont
  {Mishchenko}, \citenamefont {Mayorov}, \citenamefont {Woods}, \citenamefont
  {Wallbank} \emph {et~al.}}]{ponomarenko2013cloning}%
  \BibitemOpen
  \bibfield  {author} {\bibinfo {author} {\bibfnamefont {L.}~\bibnamefont
  {Ponomarenko}}, \bibinfo {author} {\bibfnamefont {R.}~\bibnamefont
  {Gorbachev}}, \bibinfo {author} {\bibfnamefont {G.}~\bibnamefont {Yu}},
  \bibinfo {author} {\bibfnamefont {D.}~\bibnamefont {Elias}}, \bibinfo
  {author} {\bibfnamefont {R.}~\bibnamefont {Jalil}}, \bibinfo {author}
  {\bibfnamefont {A.}~\bibnamefont {Patel}}, \bibinfo {author} {\bibfnamefont
  {A.}~\bibnamefont {Mishchenko}}, \bibinfo {author} {\bibfnamefont
  {A.}~\bibnamefont {Mayorov}}, \bibinfo {author} {\bibfnamefont
  {C.}~\bibnamefont {Woods}}, \bibinfo {author} {\bibfnamefont
  {J.}~\bibnamefont {Wallbank}}, \emph {et~al.},\ }\bibfield  {title} {\bibinfo
  {title} {Cloning of dirac fermions in graphene superlattices},\ }\href
  {https://www.nature.com/articles/nature12187} {\bibfield  {journal} {\bibinfo
   {journal} {Nature}\ }\textbf {\bibinfo {volume} {497}},\ \bibinfo {pages}
  {594} (\bibinfo {year} {2013})}\BibitemShut {NoStop}%
\bibitem [{\citenamefont {Dean}\ \emph {et~al.}(2013)\citenamefont {Dean},
  \citenamefont {Wang}, \citenamefont {Maher}, \citenamefont {Forsythe},
  \citenamefont {Ghahari}, \citenamefont {Gao}, \citenamefont {Katoch},
  \citenamefont {Ishigami}, \citenamefont {Moon}, \citenamefont {Koshino} \emph
  {et~al.}}]{dean2013hofstadter}%
  \BibitemOpen
  \bibfield  {author} {\bibinfo {author} {\bibfnamefont {C.~R.}\ \bibnamefont
  {Dean}}, \bibinfo {author} {\bibfnamefont {L.}~\bibnamefont {Wang}}, \bibinfo
  {author} {\bibfnamefont {P.}~\bibnamefont {Maher}}, \bibinfo {author}
  {\bibfnamefont {C.}~\bibnamefont {Forsythe}}, \bibinfo {author}
  {\bibfnamefont {F.}~\bibnamefont {Ghahari}}, \bibinfo {author} {\bibfnamefont
  {Y.}~\bibnamefont {Gao}}, \bibinfo {author} {\bibfnamefont {J.}~\bibnamefont
  {Katoch}}, \bibinfo {author} {\bibfnamefont {M.}~\bibnamefont {Ishigami}},
  \bibinfo {author} {\bibfnamefont {P.}~\bibnamefont {Moon}}, \bibinfo {author}
  {\bibfnamefont {M.}~\bibnamefont {Koshino}}, \emph {et~al.},\ }\bibfield
  {title} {\bibinfo {title} {Hofstadter’s butterfly and the fractal quantum
  hall effect in moir{\'e} superlattices},\ }\href
  {https://www.nature.com/articles/nature12186} {\bibfield  {journal} {\bibinfo
   {journal} {Nature}\ }\textbf {\bibinfo {volume} {497}},\ \bibinfo {pages}
  {598} (\bibinfo {year} {2013})}\BibitemShut {NoStop}%
\bibitem [{\citenamefont {Woods}\ \emph {et~al.}(2014)\citenamefont {Woods},
  \citenamefont {Britnell}, \citenamefont {Eckmann}, \citenamefont {Ma},
  \citenamefont {Lu}, \citenamefont {Guo}, \citenamefont {Lin}, \citenamefont
  {Yu}, \citenamefont {Cao}, \citenamefont {Gorbachev} \emph
  {et~al.}}]{woods2014commensurate}%
  \BibitemOpen
  \bibfield  {author} {\bibinfo {author} {\bibfnamefont {C.}~\bibnamefont
  {Woods}}, \bibinfo {author} {\bibfnamefont {L.}~\bibnamefont {Britnell}},
  \bibinfo {author} {\bibfnamefont {A.}~\bibnamefont {Eckmann}}, \bibinfo
  {author} {\bibfnamefont {R.}~\bibnamefont {Ma}}, \bibinfo {author}
  {\bibfnamefont {J.}~\bibnamefont {Lu}}, \bibinfo {author} {\bibfnamefont
  {H.}~\bibnamefont {Guo}}, \bibinfo {author} {\bibfnamefont {X.}~\bibnamefont
  {Lin}}, \bibinfo {author} {\bibfnamefont {G.}~\bibnamefont {Yu}}, \bibinfo
  {author} {\bibfnamefont {Y.}~\bibnamefont {Cao}}, \bibinfo {author}
  {\bibfnamefont {R.~V.}\ \bibnamefont {Gorbachev}}, \emph {et~al.},\
  }\bibfield  {title} {\bibinfo {title} {Commensurate--incommensurate
  transition in graphene on hexagonal boron nitride},\ }\href
  {https://www.nature.com/articles/nphys2954} {\bibfield  {journal} {\bibinfo
  {journal} {Nature physics}\ }\textbf {\bibinfo {volume} {10}},\ \bibinfo
  {pages} {451} (\bibinfo {year} {2014})}\BibitemShut {NoStop}%
\bibitem [{\citenamefont {Chen}\ \emph {et~al.}(2020)\citenamefont {Chen},
  \citenamefont {Sharpe}, \citenamefont {Fox}, \citenamefont {Zhang},
  \citenamefont {Wang}, \citenamefont {Jiang}, \citenamefont {Lyu},
  \citenamefont {Li}, \citenamefont {Watanabe}, \citenamefont {Taniguchi} \emph
  {et~al.}}]{chen2020tunable}%
  \BibitemOpen
  \bibfield  {author} {\bibinfo {author} {\bibfnamefont {G.}~\bibnamefont
  {Chen}}, \bibinfo {author} {\bibfnamefont {A.~L.}\ \bibnamefont {Sharpe}},
  \bibinfo {author} {\bibfnamefont {E.~J.}\ \bibnamefont {Fox}}, \bibinfo
  {author} {\bibfnamefont {Y.-H.}\ \bibnamefont {Zhang}}, \bibinfo {author}
  {\bibfnamefont {S.}~\bibnamefont {Wang}}, \bibinfo {author} {\bibfnamefont
  {L.}~\bibnamefont {Jiang}}, \bibinfo {author} {\bibfnamefont
  {B.}~\bibnamefont {Lyu}}, \bibinfo {author} {\bibfnamefont {H.}~\bibnamefont
  {Li}}, \bibinfo {author} {\bibfnamefont {K.}~\bibnamefont {Watanabe}},
  \bibinfo {author} {\bibfnamefont {T.}~\bibnamefont {Taniguchi}}, \emph
  {et~al.},\ }\bibfield  {title} {\bibinfo {title} {Tunable correlated chern
  insulator and ferromagnetism in a moir{\'e} superlattice},\ }\href
  {https://www.nature.com/articles/s41586-020-2049-7} {\bibfield  {journal}
  {\bibinfo  {journal} {Nature}\ }\textbf {\bibinfo {volume} {579}},\ \bibinfo
  {pages} {56} (\bibinfo {year} {2020})}\BibitemShut {NoStop}%
\bibitem [{\citenamefont {Carr}\ \emph {et~al.}(2020)\citenamefont {Carr},
  \citenamefont {Fang},\ and\ \citenamefont {Kaxiras}}]{carr2020electronic}%
  \BibitemOpen
  \bibfield  {author} {\bibinfo {author} {\bibfnamefont {S.}~\bibnamefont
  {Carr}}, \bibinfo {author} {\bibfnamefont {S.}~\bibnamefont {Fang}},\ and\
  \bibinfo {author} {\bibfnamefont {E.}~\bibnamefont {Kaxiras}},\ }\bibfield
  {title} {\bibinfo {title} {Electronic-structure methods for twisted moir{\'e}
  layers},\ }\href {https://www.nature.com/articles/s41578-020-0214-0}
  {\bibfield  {journal} {\bibinfo  {journal} {Nature Reviews Materials}\
  }\textbf {\bibinfo {volume} {5}},\ \bibinfo {pages} {748} (\bibinfo {year}
  {2020})}\BibitemShut {NoStop}%
\bibitem [{\citenamefont {{Cao}}\ \emph
  {et~al.}(2018{\natexlab{a}})\citenamefont {{Cao}}, \citenamefont {{Fatemi}},
  \citenamefont {{Fang}}, \citenamefont {{Watanabe}}, \citenamefont
  {{Taniguchi}}, \citenamefont {{Kaxiras}},\ and\ \citenamefont
  {{Jarillo}-{Herrero}}}]{cao2018unconventional}%
  \BibitemOpen
  \bibfield  {author} {\bibinfo {author} {\bibfnamefont {Y.}~\bibnamefont
  {{Cao}}}, \bibinfo {author} {\bibfnamefont {V.}~\bibnamefont {{Fatemi}}},
  \bibinfo {author} {\bibfnamefont {S.}~\bibnamefont {{Fang}}}, \bibinfo
  {author} {\bibfnamefont {K.}~\bibnamefont {{Watanabe}}}, \bibinfo {author}
  {\bibfnamefont {T.}~\bibnamefont {{Taniguchi}}}, \bibinfo {author}
  {\bibfnamefont {E.}~\bibnamefont {{Kaxiras}}},\ and\ \bibinfo {author}
  {\bibfnamefont {P.}~\bibnamefont {{Jarillo}-{Herrero}}},\ }\bibfield  {title}
  {\bibinfo {title} {{Unconventional} superconductivity in magic-angle graphene
  superlattices},\ }\href {https://www.nature.com/articles/nature26160%3C}
  {\bibfield  {journal} {\bibinfo  {journal} {{Nature}}\ }\textbf {\bibinfo
  {volume} {556}},\ \bibinfo {pages} {43} (\bibinfo {year}
  {2018}{\natexlab{a}})}\BibitemShut {NoStop}%
\bibitem [{\citenamefont {{Cao}}\ \emph
  {et~al.}(2018{\natexlab{b}})\citenamefont {{Cao}}, \citenamefont {{Fatemi}},
  \citenamefont {{Demir}}, \citenamefont {{Fang}}, \citenamefont {{Tomarken}},
  \citenamefont {{Luo}}, \citenamefont {{Sanchez}-{Yamagishi}}, \citenamefont
  {{Watanabe}}, \citenamefont {{Taniguchi}}, \citenamefont {{Kaxiras}} \emph
  {et~al.}}]{cao2018correlated}%
  \BibitemOpen
  \bibfield  {author} {\bibinfo {author} {\bibfnamefont {Y.}~\bibnamefont
  {{Cao}}}, \bibinfo {author} {\bibfnamefont {V.}~\bibnamefont {{Fatemi}}},
  \bibinfo {author} {\bibfnamefont {A.}~\bibnamefont {{Demir}}}, \bibinfo
  {author} {\bibfnamefont {S.}~\bibnamefont {{Fang}}}, \bibinfo {author}
  {\bibfnamefont {S.~L.}\ \bibnamefont {{Tomarken}}}, \bibinfo {author}
  {\bibfnamefont {J.~Y.}\ \bibnamefont {{Luo}}}, \bibinfo {author}
  {\bibfnamefont {J.~D.}\ \bibnamefont {{Sanchez}-{Yamagishi}}}, \bibinfo
  {author} {\bibfnamefont {K.}~\bibnamefont {{Watanabe}}}, \bibinfo {author}
  {\bibfnamefont {T.}~\bibnamefont {{Taniguchi}}}, \bibinfo {author}
  {\bibfnamefont {E.}~\bibnamefont {{Kaxiras}}}, \emph {et~al.},\ }\bibfield
  {title} {\bibinfo {title} {{Correlated} insulator behaviour at half-filling
  in magic-angle graphene superlattices},\ }\href
  {https://www.nature.com/articles/nature26154} {\bibfield  {journal} {\bibinfo
   {journal} {{Nature}}\ }\textbf {\bibinfo {volume} {556}},\ \bibinfo {pages}
  {80} (\bibinfo {year} {2018}{\natexlab{b}})}\BibitemShut {NoStop}%
\bibitem [{\citenamefont {{Padhi}}\ \emph {et~al.}(2018)\citenamefont
  {{Padhi}}, \citenamefont {{Setty}},\ and\ \citenamefont
  {{Phillips}}}]{padhi2018doped}%
  \BibitemOpen
  \bibfield  {author} {\bibinfo {author} {\bibfnamefont {B.}~\bibnamefont
  {{Padhi}}}, \bibinfo {author} {\bibfnamefont {C.}~\bibnamefont {{Setty}}},\
  and\ \bibinfo {author} {\bibfnamefont {P.~W.}\ \bibnamefont {{Phillips}}},\
  }\bibfield  {title} {\bibinfo {title} {{Doped} twisted bilayer graphene near
  magic angles: proximity to {Wigner} crystallization, not {Mott} insulation},\
  }\href {https://pubs.acs.org/doi/abs/10.1021/acs.nanolett.8b02033} {\bibfield
   {journal} {\bibinfo  {journal} {{Nano} letters}\ }\textbf {\bibinfo {volume}
  {18}},\ \bibinfo {pages} {6175} (\bibinfo {year} {2018})}\BibitemShut
  {NoStop}%
\bibitem [{\citenamefont {{Padhi}}\ and\ \citenamefont
  {{Phillips}}(2019)}]{padhi2019pressure}%
  \BibitemOpen
  \bibfield  {author} {\bibinfo {author} {\bibfnamefont {B.}~\bibnamefont
  {{Padhi}}}\ and\ \bibinfo {author} {\bibfnamefont {P.~W.}\ \bibnamefont
  {{Phillips}}},\ }\bibfield  {title} {\bibinfo {title} {{Pressure}-induced
  metal-insulator transition in twisted bilayer graphene},\ }\href
  {https://journals.aps.org/prb/abstract/10.1103/PhysRevB.99.205141} {\bibfield
   {journal} {\bibinfo  {journal} {{Physical} {Review} B}\ }\textbf {\bibinfo
  {volume} {99}},\ \bibinfo {pages} {205141} (\bibinfo {year}
  {2019})}\BibitemShut {NoStop}%
\bibitem [{\citenamefont {Bistritzer}\ and\ \citenamefont
  {MacDonald}(2011)}]{bistritzer2011moire}%
  \BibitemOpen
  \bibfield  {author} {\bibinfo {author} {\bibfnamefont {R.}~\bibnamefont
  {Bistritzer}}\ and\ \bibinfo {author} {\bibfnamefont {A.~H.}\ \bibnamefont
  {MacDonald}},\ }\bibfield  {title} {\bibinfo {title} {Moir{\'e} bands in
  twisted double-layer graphene},\ }\href
  {https://www.pnas.org/doi/abs/10.1073/pnas.1108174108} {\bibfield  {journal}
  {\bibinfo  {journal} {Proceedings of the National Academy of Sciences}\
  }\textbf {\bibinfo {volume} {108}},\ \bibinfo {pages} {12233} (\bibinfo
  {year} {2011})}\BibitemShut {NoStop}%
\bibitem [{\citenamefont {Koshino}\ \emph {et~al.}(2018)\citenamefont
  {Koshino}, \citenamefont {Yuan}, \citenamefont {Koretsune}, \citenamefont
  {Ochi}, \citenamefont {Kuroki},\ and\ \citenamefont
  {Fu}}]{koshino2018maximally}%
  \BibitemOpen
  \bibfield  {author} {\bibinfo {author} {\bibfnamefont {M.}~\bibnamefont
  {Koshino}}, \bibinfo {author} {\bibfnamefont {N.~F.}\ \bibnamefont {Yuan}},
  \bibinfo {author} {\bibfnamefont {T.}~\bibnamefont {Koretsune}}, \bibinfo
  {author} {\bibfnamefont {M.}~\bibnamefont {Ochi}}, \bibinfo {author}
  {\bibfnamefont {K.}~\bibnamefont {Kuroki}},\ and\ \bibinfo {author}
  {\bibfnamefont {L.}~\bibnamefont {Fu}},\ }\bibfield  {title} {\bibinfo
  {title} {Maximally localized wannier orbitals and the extended hubbard model
  for twisted bilayer graphene},\ }\href
  {https://journals.aps.org/prx/abstract/10.1103/PhysRevX.8.031087} {\bibfield
  {journal} {\bibinfo  {journal} {Physical Review X}\ }\textbf {\bibinfo
  {volume} {8}},\ \bibinfo {pages} {031087} (\bibinfo {year}
  {2018})}\BibitemShut {NoStop}%
\bibitem [{\citenamefont {Po}\ \emph {et~al.}(2018)\citenamefont {Po},
  \citenamefont {Zou}, \citenamefont {Vishwanath},\ and\ \citenamefont
  {Senthil}}]{po2018origin}%
  \BibitemOpen
  \bibfield  {author} {\bibinfo {author} {\bibfnamefont {H.~C.}\ \bibnamefont
  {Po}}, \bibinfo {author} {\bibfnamefont {L.}~\bibnamefont {Zou}}, \bibinfo
  {author} {\bibfnamefont {A.}~\bibnamefont {Vishwanath}},\ and\ \bibinfo
  {author} {\bibfnamefont {T.}~\bibnamefont {Senthil}},\ }\bibfield  {title}
  {\bibinfo {title} {Origin of mott insulating behavior and superconductivity
  in twisted bilayer graphene},\ }\href
  {https://journals.aps.org/prx/abstract/10.1103/PhysRevX.8.031089} {\bibfield
  {journal} {\bibinfo  {journal} {Physical Review X}\ }\textbf {\bibinfo
  {volume} {8}},\ \bibinfo {pages} {031089} (\bibinfo {year}
  {2018})}\BibitemShut {NoStop}%
\bibitem [{\citenamefont {Yankowitz}\ \emph
  {et~al.}(2019{\natexlab{b}})\citenamefont {Yankowitz}, \citenamefont {Chen},
  \citenamefont {Polshyn}, \citenamefont {Zhang}, \citenamefont {Watanabe},
  \citenamefont {Taniguchi}, \citenamefont {Graf}, \citenamefont {Young},\ and\
  \citenamefont {Dean}}]{yankowitz2019tuning}%
  \BibitemOpen
  \bibfield  {author} {\bibinfo {author} {\bibfnamefont {M.}~\bibnamefont
  {Yankowitz}}, \bibinfo {author} {\bibfnamefont {S.}~\bibnamefont {Chen}},
  \bibinfo {author} {\bibfnamefont {H.}~\bibnamefont {Polshyn}}, \bibinfo
  {author} {\bibfnamefont {Y.}~\bibnamefont {Zhang}}, \bibinfo {author}
  {\bibfnamefont {K.}~\bibnamefont {Watanabe}}, \bibinfo {author}
  {\bibfnamefont {T.}~\bibnamefont {Taniguchi}}, \bibinfo {author}
  {\bibfnamefont {D.}~\bibnamefont {Graf}}, \bibinfo {author} {\bibfnamefont
  {A.~F.}\ \bibnamefont {Young}},\ and\ \bibinfo {author} {\bibfnamefont
  {C.~R.}\ \bibnamefont {Dean}},\ }\bibfield  {title} {\bibinfo {title} {Tuning
  superconductivity in twisted bilayer graphene},\ }\href
  {https://doi.org/10.1126/science.aav1910} {\bibfield  {journal} {\bibinfo
  {journal} {Science}\ }\textbf {\bibinfo {volume} {363}},\ \bibinfo {pages}
  {1059} (\bibinfo {year} {2019}{\natexlab{b}})}\BibitemShut {NoStop}%
\bibitem [{\citenamefont {{Long}}\ \emph {et~al.}(2022)\citenamefont {{Long}},
  \citenamefont {{Pantale}{\'o}n}, \citenamefont {{Zhan}}, \citenamefont
  {{Guinea}}, \citenamefont {{Silva}-{Guill}{\'e}n},\ and\ \citenamefont
  {{Yuan}}}]{long2022atomistic}%
  \BibitemOpen
  \bibfield  {author} {\bibinfo {author} {\bibfnamefont {M.}~\bibnamefont
  {{Long}}}, \bibinfo {author} {\bibfnamefont {P.~A.}\ \bibnamefont
  {{Pantale}{\'o}n}}, \bibinfo {author} {\bibfnamefont {Z.}~\bibnamefont
  {{Zhan}}}, \bibinfo {author} {\bibfnamefont {F.}~\bibnamefont {{Guinea}}},
  \bibinfo {author} {\bibfnamefont {J.~{\'A}.}\ \bibnamefont
  {{Silva}-{Guill}{\'e}n}},\ and\ \bibinfo {author} {\bibfnamefont
  {S.}~\bibnamefont {{Yuan}}},\ }\bibfield  {title} {\bibinfo {title} {{An}
  atomistic approach for the structural and electronic properties of twisted
  bilayer graphene-boron nitride heterostructures},\ }\href
  {https://www.nature.com/articles/s41524-022-00763-1} {\bibfield  {journal}
  {\bibinfo  {journal} {npj {Computational} {Materials}}\ }\textbf {\bibinfo
  {volume} {8}},\ \bibinfo {pages} {73} (\bibinfo {year} {2022})}\BibitemShut
  {NoStop}%
\bibitem [{\citenamefont {Reina}\ \emph {et~al.}(2009)\citenamefont {Reina},
  \citenamefont {Jia}, \citenamefont {Ho}, \citenamefont {Nezich},
  \citenamefont {Son}, \citenamefont {Bulovic}, \citenamefont {Dresselhaus},\
  and\ \citenamefont {Kong}}]{reina2009large}%
  \BibitemOpen
  \bibfield  {author} {\bibinfo {author} {\bibfnamefont {A.}~\bibnamefont
  {Reina}}, \bibinfo {author} {\bibfnamefont {X.}~\bibnamefont {Jia}}, \bibinfo
  {author} {\bibfnamefont {J.}~\bibnamefont {Ho}}, \bibinfo {author}
  {\bibfnamefont {D.}~\bibnamefont {Nezich}}, \bibinfo {author} {\bibfnamefont
  {H.}~\bibnamefont {Son}}, \bibinfo {author} {\bibfnamefont {V.}~\bibnamefont
  {Bulovic}}, \bibinfo {author} {\bibfnamefont {M.~S.}\ \bibnamefont
  {Dresselhaus}},\ and\ \bibinfo {author} {\bibfnamefont {J.}~\bibnamefont
  {Kong}},\ }\bibfield  {title} {\bibinfo {title} {Large area, few-layer
  graphene films on arbitrary substrates by chemical vapor deposition},\ }\href
  {https://doi.org/10.1021/nl801827v} {\bibfield  {journal} {\bibinfo
  {journal} {Nano letters}\ }\textbf {\bibinfo {volume} {9}},\ \bibinfo {pages}
  {30} (\bibinfo {year} {2009})}\BibitemShut {NoStop}%
\bibitem [{\citenamefont {Rong}\ and\ \citenamefont
  {Kuiper}(1993)}]{rong1993electronic}%
  \BibitemOpen
  \bibfield  {author} {\bibinfo {author} {\bibfnamefont {Z.~Y.}\ \bibnamefont
  {Rong}}\ and\ \bibinfo {author} {\bibfnamefont {P.}~\bibnamefont {Kuiper}},\
  }\bibfield  {title} {\bibinfo {title} {Electronic effects in scanning
  tunneling microscopy: Moir{\'e} pattern on a graphite surface},\ }\href
  {https://doi.org/10.1103/PhysRevB.48.17427} {\bibfield  {journal} {\bibinfo
  {journal} {Physical Review B}\ }\textbf {\bibinfo {volume} {48}},\ \bibinfo
  {pages} {17427} (\bibinfo {year} {1993})}\BibitemShut {NoStop}%
\bibitem [{\citenamefont {Zhou}\ \emph {et~al.}(2006)\citenamefont {Zhou},
  \citenamefont {Gweon},\ and\ \citenamefont {Lanzara}}]{zhou2006low}%
  \BibitemOpen
  \bibfield  {author} {\bibinfo {author} {\bibfnamefont {S.}~\bibnamefont
  {Zhou}}, \bibinfo {author} {\bibfnamefont {G.-H.}\ \bibnamefont {Gweon}},\
  and\ \bibinfo {author} {\bibfnamefont {A.}~\bibnamefont {Lanzara}},\
  }\bibfield  {title} {\bibinfo {title} {Low energy excitations in graphite:
  the role of dimensionality and lattice defects},\ }\href
  {https://doi.org/10.1016/j.aop.2006.04.011} {\bibfield  {journal} {\bibinfo
  {journal} {Annals of Physics}\ }\textbf {\bibinfo {volume} {321}},\ \bibinfo
  {pages} {1730} (\bibinfo {year} {2006})}\BibitemShut {NoStop}%
\bibitem [{\citenamefont {Peres}\ \emph {et~al.}(2006)\citenamefont {Peres},
  \citenamefont {Guinea},\ and\ \citenamefont {Neto}}]{peres2006electronic}%
  \BibitemOpen
  \bibfield  {author} {\bibinfo {author} {\bibfnamefont {N.}~\bibnamefont
  {Peres}}, \bibinfo {author} {\bibfnamefont {F.}~\bibnamefont {Guinea}},\ and\
  \bibinfo {author} {\bibfnamefont {A.~C.}\ \bibnamefont {Neto}},\ }\bibfield
  {title} {\bibinfo {title} {Electronic properties of disordered
  two-dimensional carbon},\ }\href {https://doi.org/10.1103/PhysRevB.73.125411}
  {\bibfield  {journal} {\bibinfo  {journal} {Physical Review B}\ }\textbf
  {\bibinfo {volume} {73}},\ \bibinfo {pages} {125411} (\bibinfo {year}
  {2006})}\BibitemShut {NoStop}%
\bibitem [{\citenamefont {Shima}\ and\ \citenamefont
  {Aoki}(1993)}]{shima1993electronic}%
  \BibitemOpen
  \bibfield  {author} {\bibinfo {author} {\bibfnamefont {N.}~\bibnamefont
  {Shima}}\ and\ \bibinfo {author} {\bibfnamefont {H.}~\bibnamefont {Aoki}},\
  }\bibfield  {title} {\bibinfo {title} {Electronic structure of
  super-honeycomb systems: A peculiar realization of semimetal/semiconductor
  classes and ferromagnetism},\ }\href
  {https://doi.org/10.1103/PhysRevLett.71.4389} {\bibfield  {journal} {\bibinfo
   {journal} {Physical review letters}\ }\textbf {\bibinfo {volume} {71}},\
  \bibinfo {pages} {4389} (\bibinfo {year} {1993})}\BibitemShut {NoStop}%
\bibitem [{\citenamefont {Nguyen}\ \emph {et~al.}(2014)\citenamefont {Nguyen},
  \citenamefont {Lee}, \citenamefont {Yoon},\ and\ \citenamefont
  {Cheong}}]{nguyen2014excitation}%
  \BibitemOpen
  \bibfield  {author} {\bibinfo {author} {\bibfnamefont {T.~A.}\ \bibnamefont
  {Nguyen}}, \bibinfo {author} {\bibfnamefont {J.-U.}\ \bibnamefont {Lee}},
  \bibinfo {author} {\bibfnamefont {D.}~\bibnamefont {Yoon}},\ and\ \bibinfo
  {author} {\bibfnamefont {H.}~\bibnamefont {Cheong}},\ }\bibfield  {title}
  {\bibinfo {title} {Excitation energy dependent raman signatures of aba-and
  abc-stacked few-layer graphene},\ }\href {https://doi.org/10.1038/srep04630}
  {\bibfield  {journal} {\bibinfo  {journal} {Scientific reports}\ }\textbf
  {\bibinfo {volume} {4}},\ \bibinfo {pages} {1} (\bibinfo {year}
  {2014})}\BibitemShut {NoStop}%
\bibitem [{\citenamefont {Can}\ \emph {et~al.}(2021)\citenamefont {Can},
  \citenamefont {Tummuru}, \citenamefont {Day}, \citenamefont {Elfimov},
  \citenamefont {Damascelli},\ and\ \citenamefont {Franz}}]{can2021high}%
  \BibitemOpen
  \bibfield  {author} {\bibinfo {author} {\bibfnamefont {O.}~\bibnamefont
  {Can}}, \bibinfo {author} {\bibfnamefont {T.}~\bibnamefont {Tummuru}},
  \bibinfo {author} {\bibfnamefont {R.~P.}\ \bibnamefont {Day}}, \bibinfo
  {author} {\bibfnamefont {I.}~\bibnamefont {Elfimov}}, \bibinfo {author}
  {\bibfnamefont {A.}~\bibnamefont {Damascelli}},\ and\ \bibinfo {author}
  {\bibfnamefont {M.}~\bibnamefont {Franz}},\ }\bibfield  {title} {\bibinfo
  {title} {High-temperature topological superconductivity in twisted
  double-layer copper oxides},\ }\href
  {https://www.nature.com/articles/s41567-020-01142-7} {\bibfield  {journal}
  {\bibinfo  {journal} {Nature Physics}\ }\textbf {\bibinfo {volume} {17}},\
  \bibinfo {pages} {519} (\bibinfo {year} {2021})}\BibitemShut {NoStop}%
\bibitem [{\citenamefont {Peng}\ \emph {et~al.}(2022)\citenamefont {Peng},
  \citenamefont {Guo}, \citenamefont {Xiao}, \citenamefont {Li}, \citenamefont
  {Strempfer}, \citenamefont {Choi}, \citenamefont {Yan}, \citenamefont {Luo},
  \citenamefont {Huang}, \citenamefont {Jia} \emph
  {et~al.}}]{peng2022observation}%
  \BibitemOpen
  \bibfield  {author} {\bibinfo {author} {\bibfnamefont {Y.}~\bibnamefont
  {Peng}}, \bibinfo {author} {\bibfnamefont {X.}~\bibnamefont {Guo}}, \bibinfo
  {author} {\bibfnamefont {Q.}~\bibnamefont {Xiao}}, \bibinfo {author}
  {\bibfnamefont {Q.}~\bibnamefont {Li}}, \bibinfo {author} {\bibfnamefont
  {J.}~\bibnamefont {Strempfer}}, \bibinfo {author} {\bibfnamefont
  {Y.}~\bibnamefont {Choi}}, \bibinfo {author} {\bibfnamefont {D.}~\bibnamefont
  {Yan}}, \bibinfo {author} {\bibfnamefont {H.}~\bibnamefont {Luo}}, \bibinfo
  {author} {\bibfnamefont {Y.}~\bibnamefont {Huang}}, \bibinfo {author}
  {\bibfnamefont {S.}~\bibnamefont {Jia}}, \emph {et~al.},\ }\bibfield  {title}
  {\bibinfo {title} {Observation of orbital order in the van der waals material
  1 t-tise 2},\ }\href
  {https://journals.aps.org/prresearch/abstract/10.1103/PhysRevResearch.4.033053}
  {\bibfield  {journal} {\bibinfo  {journal} {Physical Review Research}\
  }\textbf {\bibinfo {volume} {4}},\ \bibinfo {pages} {033053} (\bibinfo {year}
  {2022})}\BibitemShut {NoStop}%
\bibitem [{\citenamefont {Peng}\ \emph {et~al.}(2024)\citenamefont {Peng},
  \citenamefont {Boukahil}, \citenamefont {Krongchon}, \citenamefont {Xiao},
  \citenamefont {Husain}, \citenamefont {Lee}, \citenamefont {Li},
  \citenamefont {Alatas}, \citenamefont {Said}, \citenamefont {Yan} \emph
  {et~al.}}]{peng2024observation}%
  \BibitemOpen
  \bibfield  {author} {\bibinfo {author} {\bibfnamefont {Y.}~\bibnamefont
  {Peng}}, \bibinfo {author} {\bibfnamefont {I.}~\bibnamefont {Boukahil}},
  \bibinfo {author} {\bibfnamefont {K.}~\bibnamefont {Krongchon}}, \bibinfo
  {author} {\bibfnamefont {Q.}~\bibnamefont {Xiao}}, \bibinfo {author}
  {\bibfnamefont {A.}~\bibnamefont {Husain}}, \bibinfo {author} {\bibfnamefont
  {S.}~\bibnamefont {Lee}}, \bibinfo {author} {\bibfnamefont {Q.}~\bibnamefont
  {Li}}, \bibinfo {author} {\bibfnamefont {A.}~\bibnamefont {Alatas}}, \bibinfo
  {author} {\bibfnamefont {A.}~\bibnamefont {Said}}, \bibinfo {author}
  {\bibfnamefont {H.}~\bibnamefont {Yan}}, \emph {et~al.},\ }\bibfield  {title}
  {\bibinfo {title} {Observation of van der waals phonons in the single-layer
  cuprate (bi, pb) 2 (sr, la) 2 cuo 6+ $\delta$},\ }\href
  {https://journals.aps.org/prmaterials/abstract/10.1103/PhysRevMaterials.8.024804}
  {\bibfield  {journal} {\bibinfo  {journal} {Physical Review Materials}\
  }\textbf {\bibinfo {volume} {8}},\ \bibinfo {pages} {024804} (\bibinfo {year}
  {2024})}\BibitemShut {NoStop}%
\bibitem [{\citenamefont {Zhou}\ \emph {et~al.}(2015)\citenamefont {Zhou},
  \citenamefont {Han}, \citenamefont {Dai}, \citenamefont {Sun},\ and\
  \citenamefont {Srolovitz}}]{zhou2015van}%
  \BibitemOpen
  \bibfield  {author} {\bibinfo {author} {\bibfnamefont {S.}~\bibnamefont
  {Zhou}}, \bibinfo {author} {\bibfnamefont {J.}~\bibnamefont {Han}}, \bibinfo
  {author} {\bibfnamefont {S.}~\bibnamefont {Dai}}, \bibinfo {author}
  {\bibfnamefont {J.}~\bibnamefont {Sun}},\ and\ \bibinfo {author}
  {\bibfnamefont {D.~J.}\ \bibnamefont {Srolovitz}},\ }\bibfield  {title}
  {\bibinfo {title} {van der waals bilayer energetics: Generalized
  stacking-fault energy of graphene, boron nitride, and graphene/boron nitride
  bilayers},\ }\href
  {https://journals.aps.org/prb/abstract/10.1103/PhysRevB.92.155438} {\bibfield
   {journal} {\bibinfo  {journal} {Physical Review B}\ }\textbf {\bibinfo
  {volume} {92}},\ \bibinfo {pages} {155438} (\bibinfo {year}
  {2015})}\BibitemShut {NoStop}%
\bibitem [{\citenamefont {Zhang}\ \emph {et~al.}(2024)\citenamefont {Zhang},
  \citenamefont {Qiu}, \citenamefont {Liao}, \citenamefont {He},\ and\
  \citenamefont {Ni}}]{zhang2024impact}%
  \BibitemOpen
  \bibfield  {author} {\bibinfo {author} {\bibfnamefont {B.}~\bibnamefont
  {Zhang}}, \bibinfo {author} {\bibfnamefont {W.}~\bibnamefont {Qiu}}, \bibinfo
  {author} {\bibfnamefont {X.}~\bibnamefont {Liao}}, \bibinfo {author}
  {\bibfnamefont {L.}~\bibnamefont {He}},\ and\ \bibinfo {author}
  {\bibfnamefont {Y.}~\bibnamefont {Ni}},\ }\bibfield  {title} {\bibinfo
  {title} {Impact of out-of-plane deformation on atomic reconstruction in
  twisted van der waals bilayers},\ }\href
  {https://doi.org/10.1016/j.jmps.2024.105693} {\bibfield  {journal} {\bibinfo
  {journal} {Journal of the Mechanics and Physics of Solids}\ ,\ \bibinfo
  {pages} {105693}} (\bibinfo {year} {2024})}\BibitemShut {NoStop}%
\bibitem [{\citenamefont {Dubecky}\ \emph {et~al.}(2016)\citenamefont
  {Dubecky}, \citenamefont {Mitas},\ and\ \citenamefont
  {Jurecka}}]{dubecky2016noncovalent}%
  \BibitemOpen
  \bibfield  {author} {\bibinfo {author} {\bibfnamefont {M.}~\bibnamefont
  {Dubecky}}, \bibinfo {author} {\bibfnamefont {L.}~\bibnamefont {Mitas}},\
  and\ \bibinfo {author} {\bibfnamefont {P.}~\bibnamefont {Jurecka}},\
  }\bibfield  {title} {\bibinfo {title} {Noncovalent interactions by quantum
  monte carlo},\ }\href {https://doi.org/10.1021/acs.chemrev.5b00577}
  {\bibfield  {journal} {\bibinfo  {journal} {Chemical Reviews}\ }\textbf
  {\bibinfo {volume} {116}},\ \bibinfo {pages} {5188} (\bibinfo {year}
  {2016})}\BibitemShut {NoStop}%
\bibitem [{\citenamefont {Krongchon}\ \emph {et~al.}(2017)\citenamefont
  {Krongchon}, \citenamefont {Busemeyer},\ and\ \citenamefont
  {Wagner}}]{krongchon2017accurate}%
  \BibitemOpen
  \bibfield  {author} {\bibinfo {author} {\bibfnamefont {K.}~\bibnamefont
  {Krongchon}}, \bibinfo {author} {\bibfnamefont {B.}~\bibnamefont
  {Busemeyer}},\ and\ \bibinfo {author} {\bibfnamefont {L.~K.}\ \bibnamefont
  {Wagner}},\ }\bibfield  {title} {\bibinfo {title} {Accurate barrier heights
  using diffusion monte carlo},\ }\href {https://doi.org/10.1063/1.4979059}
  {\bibfield  {journal} {\bibinfo  {journal} {The Journal of Chemical Physics}\
  }\textbf {\bibinfo {volume} {146}} (\bibinfo {year} {2017})}\BibitemShut
  {NoStop}%
\bibitem [{\citenamefont {Williams}\ \emph {et~al.}(2020)\citenamefont
  {Williams}, \citenamefont {Yao}, \citenamefont {Li}, \citenamefont {Chen},
  \citenamefont {Shi}, \citenamefont {Motta}, \citenamefont {Niu},
  \citenamefont {Ray}, \citenamefont {Guo}, \citenamefont {Anderson} \emph
  {et~al.}}]{williams2020direct}%
  \BibitemOpen
  \bibfield  {author} {\bibinfo {author} {\bibfnamefont {K.~T.}\ \bibnamefont
  {Williams}}, \bibinfo {author} {\bibfnamefont {Y.}~\bibnamefont {Yao}},
  \bibinfo {author} {\bibfnamefont {J.}~\bibnamefont {Li}}, \bibinfo {author}
  {\bibfnamefont {L.}~\bibnamefont {Chen}}, \bibinfo {author} {\bibfnamefont
  {H.}~\bibnamefont {Shi}}, \bibinfo {author} {\bibfnamefont {M.}~\bibnamefont
  {Motta}}, \bibinfo {author} {\bibfnamefont {C.}~\bibnamefont {Niu}}, \bibinfo
  {author} {\bibfnamefont {U.}~\bibnamefont {Ray}}, \bibinfo {author}
  {\bibfnamefont {S.}~\bibnamefont {Guo}}, \bibinfo {author} {\bibfnamefont
  {R.~J.}\ \bibnamefont {Anderson}}, \emph {et~al.},\ }\bibfield  {title}
  {\bibinfo {title} {Direct comparison of many-body methods for realistic
  electronic hamiltonians},\ }\href
  {https://journals.aps.org/prx/abstract/10.1103/PhysRevX.10.011041} {\bibfield
   {journal} {\bibinfo  {journal} {Physical Review X}\ }\textbf {\bibinfo
  {volume} {10}},\ \bibinfo {pages} {011041} (\bibinfo {year}
  {2020})}\BibitemShut {NoStop}%
\bibitem [{\citenamefont {Giovannetti}\ \emph {et~al.}(2007)\citenamefont
  {Giovannetti}, \citenamefont {Khomyakov}, \citenamefont {Brocks},
  \citenamefont {Kelly},\ and\ \citenamefont {Van
  Den~Brink}}]{giovannetti2007substrate}%
  \BibitemOpen
  \bibfield  {author} {\bibinfo {author} {\bibfnamefont {G.}~\bibnamefont
  {Giovannetti}}, \bibinfo {author} {\bibfnamefont {P.~A.}\ \bibnamefont
  {Khomyakov}}, \bibinfo {author} {\bibfnamefont {G.}~\bibnamefont {Brocks}},
  \bibinfo {author} {\bibfnamefont {P.~J.}\ \bibnamefont {Kelly}},\ and\
  \bibinfo {author} {\bibfnamefont {J.}~\bibnamefont {Van Den~Brink}},\
  }\bibfield  {title} {\bibinfo {title} {Substrate-induced band gap in graphene
  on hexagonal boron nitride: Ab initio density functional calculations},\
  }\href {https://journals.aps.org/prb/abstract/10.1103/PhysRevB.76.073103}
  {\bibfield  {journal} {\bibinfo  {journal} {Physical Review B—Condensed
  Matter and Materials Physics}\ }\textbf {\bibinfo {volume} {76}},\ \bibinfo
  {pages} {073103} (\bibinfo {year} {2007})}\BibitemShut {NoStop}%
\bibitem [{\citenamefont {{Correa}}\ \emph {et~al.}(2014)\citenamefont
  {{Correa}}, \citenamefont {{Pacheco}},\ and\ \citenamefont
  {{Morell}}}]{correa2014optical}%
  \BibitemOpen
  \bibfield  {author} {\bibinfo {author} {\bibfnamefont {J.~D.}\ \bibnamefont
  {{Correa}}}, \bibinfo {author} {\bibfnamefont {M.}~\bibnamefont
  {{Pacheco}}},\ and\ \bibinfo {author} {\bibfnamefont {E.~S.}\ \bibnamefont
  {{Morell}}},\ }\bibfield  {title} {\bibinfo {title} {{Optical} absorption
  spectrum of rotated trilayer graphene},\ }\href
  {https://link.springer.com/article/10.1007/s10853-013-7744-4} {\bibfield
  {journal} {\bibinfo  {journal} {{Journal} of {Materials} {Science}}\ }\textbf
  {\bibinfo {volume} {49}},\ \bibinfo {pages} {642} (\bibinfo {year}
  {2014})}\BibitemShut {NoStop}%
\bibitem [{\citenamefont {Sevilla}\ and\ \citenamefont
  {Putungan}(2021)}]{sevilla2021graphene}%
  \BibitemOpen
  \bibfield  {author} {\bibinfo {author} {\bibfnamefont {J.~R.~M.}\
  \bibnamefont {Sevilla}}\ and\ \bibinfo {author} {\bibfnamefont {D.~B.}\
  \bibnamefont {Putungan}},\ }\bibfield  {title} {\bibinfo {title}
  {Graphene-hexagonal boron nitride van der waals heterostructures: an
  examination of the effects of different van der waals corrections},\ }\href
  {https://iopscience.iop.org/article/10.1088/2053-1591/ac187d/meta} {\bibfield
   {journal} {\bibinfo  {journal} {Materials Research Express}\ }\textbf
  {\bibinfo {volume} {8}},\ \bibinfo {pages} {085601} (\bibinfo {year}
  {2021})}\BibitemShut {NoStop}%
\bibitem [{\citenamefont {{Mostaani}}\ \emph {et~al.}(2015)\citenamefont
  {{Mostaani}}, \citenamefont {{Drummond}},\ and\ \citenamefont
  {{Fal}’{Ko}}}]{mostaani2015quantum}%
  \BibitemOpen
  \bibfield  {author} {\bibinfo {author} {\bibfnamefont {E.}~\bibnamefont
  {{Mostaani}}}, \bibinfo {author} {\bibfnamefont {N.}~\bibnamefont
  {{Drummond}}},\ and\ \bibinfo {author} {\bibfnamefont {V.}~\bibnamefont
  {{Fal}’{Ko}}},\ }\bibfield  {title} {\bibinfo {title} {{Quantum} {Monte}
  {Carlo} calculation of the binding energy of bilayer graphene},\ }\href
  {https://journals.aps.org/prl/abstract/10.1103/PhysRevLett.115.115501}
  {\bibfield  {journal} {\bibinfo  {journal} {{Physical} {Review} {Letters}}\
  }\textbf {\bibinfo {volume} {115}},\ \bibinfo {pages} {115501} (\bibinfo
  {year} {2015})}\BibitemShut {NoStop}%
\bibitem [{\citenamefont {{Krongchon}}\ \emph {et~al.}(2023)\citenamefont
  {{Krongchon}}, \citenamefont {{Rakib}}, \citenamefont {{Pathak}},
  \citenamefont {{Ertekin}}, \citenamefont {{Johnson}},\ and\ \citenamefont
  {{Wagner}}}]{krongchon2023registry}%
  \BibitemOpen
  \bibfield  {author} {\bibinfo {author} {\bibfnamefont {K.}~\bibnamefont
  {{Krongchon}}}, \bibinfo {author} {\bibfnamefont {T.}~\bibnamefont
  {{Rakib}}}, \bibinfo {author} {\bibfnamefont {S.}~\bibnamefont {{Pathak}}},
  \bibinfo {author} {\bibfnamefont {E.}~\bibnamefont {{Ertekin}}}, \bibinfo
  {author} {\bibfnamefont {H.~T.}\ \bibnamefont {{Johnson}}},\ and\ \bibinfo
  {author} {\bibfnamefont {L.~K.}\ \bibnamefont {{Wagner}}},\ }\bibfield
  {title} {\bibinfo {title} {{Registry}-dependent potential energy and lattice
  corrugation of twisted bilayer graphene from quantum {Monte} {Carlo}},\
  }\href {https://doi.org/10.1103/PhysRevB.108.235403} {\bibfield  {journal}
  {\bibinfo  {journal} {{Physical} {Review} B}\ }\textbf {\bibinfo {volume}
  {108}},\ \bibinfo {pages} {235403} (\bibinfo {year} {2023})}\BibitemShut
  {NoStop}%
\bibitem [{\citenamefont {Ambrosetti}\ \emph {et~al.}(2014)\citenamefont
  {Ambrosetti}, \citenamefont {Reilly}, \citenamefont {DiStasio},\ and\
  \citenamefont {Tkatchenko}}]{ambrosetti2014long}%
  \BibitemOpen
  \bibfield  {author} {\bibinfo {author} {\bibfnamefont {A.}~\bibnamefont
  {Ambrosetti}}, \bibinfo {author} {\bibfnamefont {A.~M.}\ \bibnamefont
  {Reilly}}, \bibinfo {author} {\bibfnamefont {R.~A.}\ \bibnamefont
  {DiStasio}},\ and\ \bibinfo {author} {\bibfnamefont {A.}~\bibnamefont
  {Tkatchenko}},\ }\bibfield  {title} {\bibinfo {title} {Long-range correlation
  energy calculated from coupled atomic response functions},\ }\href
  {https://doi.org/10.1063/1.4865104} {\bibfield  {journal} {\bibinfo
  {journal} {The Journal of chemical physics}\ }\textbf {\bibinfo {volume}
  {140}} (\bibinfo {year} {2014})}\BibitemShut {NoStop}%
\bibitem [{\citenamefont {Gurtubay}\ and\ \citenamefont
  {Needs}(2007)}]{gurtubay2007dissociation}%
  \BibitemOpen
  \bibfield  {author} {\bibinfo {author} {\bibfnamefont {I.}~\bibnamefont
  {Gurtubay}}\ and\ \bibinfo {author} {\bibfnamefont {R.}~\bibnamefont
  {Needs}},\ }\bibfield  {title} {\bibinfo {title} {Dissociation energy of the
  water dimer from quantum monte carlo calculations},\ }\href
  {https://pubs.aip.org/aip/jcp/article/127/12/124306/927426} {\bibfield
  {journal} {\bibinfo  {journal} {The Journal of chemical physics}\ }\textbf
  {\bibinfo {volume} {127}} (\bibinfo {year} {2007})}\BibitemShut {NoStop}%
\bibitem [{\citenamefont {Shulenburger}\ \emph {et~al.}(2015)\citenamefont
  {Shulenburger}, \citenamefont {Baczewski}, \citenamefont {Zhu}, \citenamefont
  {Guan},\ and\ \citenamefont {Tomanek}}]{shulenburger2015nature}%
  \BibitemOpen
  \bibfield  {author} {\bibinfo {author} {\bibfnamefont {L.}~\bibnamefont
  {Shulenburger}}, \bibinfo {author} {\bibfnamefont {A.~D.}\ \bibnamefont
  {Baczewski}}, \bibinfo {author} {\bibfnamefont {Z.}~\bibnamefont {Zhu}},
  \bibinfo {author} {\bibfnamefont {J.}~\bibnamefont {Guan}},\ and\ \bibinfo
  {author} {\bibfnamefont {D.}~\bibnamefont {Tomanek}},\ }\bibfield  {title}
  {\bibinfo {title} {The nature of the interlayer interaction in bulk and
  few-layer phosphorus},\ }\href
  {https://pubs.acs.org/doi/abs/10.1021/acs.nanolett.5b03615} {\bibfield
  {journal} {\bibinfo  {journal} {Nano letters}\ }\textbf {\bibinfo {volume}
  {15}},\ \bibinfo {pages} {8170} (\bibinfo {year} {2015})}\BibitemShut
  {NoStop}%
\bibitem [{\citenamefont {Wu}\ \emph {et~al.}(2016)\citenamefont {Wu},
  \citenamefont {Wagner},\ and\ \citenamefont {Aluru}}]{wu2016hexagonal}%
  \BibitemOpen
  \bibfield  {author} {\bibinfo {author} {\bibfnamefont {Y.}~\bibnamefont
  {Wu}}, \bibinfo {author} {\bibfnamefont {L.~K.}\ \bibnamefont {Wagner}},\
  and\ \bibinfo {author} {\bibfnamefont {N.~R.}\ \bibnamefont {Aluru}},\
  }\bibfield  {title} {\bibinfo {title} {Hexagonal boron nitride and water
  interaction parameters},\ }\href
  {https://pubs.aip.org/aip/jcp/article/144/16/164118/961544} {\bibfield
  {journal} {\bibinfo  {journal} {The Journal of chemical physics}\ }\textbf
  {\bibinfo {volume} {144}} (\bibinfo {year} {2016})}\BibitemShut {NoStop}%
\bibitem [{\citenamefont {Kadioglu}\ \emph {et~al.}(2018)\citenamefont
  {Kadioglu}, \citenamefont {Santana}, \citenamefont {{\"O}zaydin},
  \citenamefont {Ersan}, \citenamefont {Akt{\"u}rk}, \citenamefont
  {Akt{\"u}rk},\ and\ \citenamefont {Reboredo}}]{kadioglu2018diffusion}%
  \BibitemOpen
  \bibfield  {author} {\bibinfo {author} {\bibfnamefont {Y.}~\bibnamefont
  {Kadioglu}}, \bibinfo {author} {\bibfnamefont {J.~A.}\ \bibnamefont
  {Santana}}, \bibinfo {author} {\bibfnamefont {H.~D.}\ \bibnamefont
  {{\"O}zaydin}}, \bibinfo {author} {\bibfnamefont {F.}~\bibnamefont {Ersan}},
  \bibinfo {author} {\bibfnamefont {O.~{\"U}.}\ \bibnamefont {Akt{\"u}rk}},
  \bibinfo {author} {\bibfnamefont {E.}~\bibnamefont {Akt{\"u}rk}},\ and\
  \bibinfo {author} {\bibfnamefont {F.~A.}\ \bibnamefont {Reboredo}},\
  }\bibfield  {title} {\bibinfo {title} {Diffusion quantum monte carlo and
  density functional calculations of the structural stability of bilayer
  arsenene},\ }\href
  {https://pubs.aip.org/aip/jcp/article/148/21/214706/1079303} {\bibfield
  {journal} {\bibinfo  {journal} {The Journal of chemical physics}\ }\textbf
  {\bibinfo {volume} {148}} (\bibinfo {year} {2018})}\BibitemShut {NoStop}%
\bibitem [{\citenamefont {Ahn}\ \emph {et~al.}(2021)\citenamefont {Ahn},
  \citenamefont {Hong}, \citenamefont {Lee}, \citenamefont {Shin},
  \citenamefont {Benali},\ and\ \citenamefont {Kwon}}]{ahn2021metastable}%
  \BibitemOpen
  \bibfield  {author} {\bibinfo {author} {\bibfnamefont {J.}~\bibnamefont
  {Ahn}}, \bibinfo {author} {\bibfnamefont {I.}~\bibnamefont {Hong}}, \bibinfo
  {author} {\bibfnamefont {G.}~\bibnamefont {Lee}}, \bibinfo {author}
  {\bibfnamefont {H.}~\bibnamefont {Shin}}, \bibinfo {author} {\bibfnamefont
  {A.}~\bibnamefont {Benali}},\ and\ \bibinfo {author} {\bibfnamefont
  {Y.}~\bibnamefont {Kwon}},\ }\bibfield  {title} {\bibinfo {title} {Metastable
  metallic phase of a bilayer blue phosphorene induced by interlayer bonding
  and intralayer charge redistributions},\ }\href
  {https://doi.org/10.1021/acs.jpclett.1c03045} {\bibfield  {journal} {\bibinfo
   {journal} {The Journal of Physical Chemistry Letters}\ }\textbf {\bibinfo
  {volume} {12}},\ \bibinfo {pages} {10981} (\bibinfo {year}
  {2021})}\BibitemShut {NoStop}%
\bibitem [{\citenamefont {Staros}\ \emph {et~al.}(2022)\citenamefont {Staros},
  \citenamefont {Hu}, \citenamefont {Tiihonen}, \citenamefont {Nanguneri},
  \citenamefont {Krogel}, \citenamefont {Bennett}, \citenamefont {Heinonen},
  \citenamefont {Ganesh},\ and\ \citenamefont
  {Rubenstein}}]{staros2022combined}%
  \BibitemOpen
  \bibfield  {author} {\bibinfo {author} {\bibfnamefont {D.}~\bibnamefont
  {Staros}}, \bibinfo {author} {\bibfnamefont {G.}~\bibnamefont {Hu}}, \bibinfo
  {author} {\bibfnamefont {J.}~\bibnamefont {Tiihonen}}, \bibinfo {author}
  {\bibfnamefont {R.}~\bibnamefont {Nanguneri}}, \bibinfo {author}
  {\bibfnamefont {J.}~\bibnamefont {Krogel}}, \bibinfo {author} {\bibfnamefont
  {M.~C.}\ \bibnamefont {Bennett}}, \bibinfo {author} {\bibfnamefont
  {O.}~\bibnamefont {Heinonen}}, \bibinfo {author} {\bibfnamefont
  {P.}~\bibnamefont {Ganesh}},\ and\ \bibinfo {author} {\bibfnamefont
  {B.}~\bibnamefont {Rubenstein}},\ }\bibfield  {title} {\bibinfo {title} {A
  combined first principles study of the structural, magnetic, and phonon
  properties of monolayer cri3},\ }\href
  {https://pubs.aip.org/aip/jcp/article/156/1/014707/2840559} {\bibfield
  {journal} {\bibinfo  {journal} {The Journal of Chemical Physics}\ }\textbf
  {\bibinfo {volume} {156}} (\bibinfo {year} {2022})}\BibitemShut {NoStop}%
\bibitem [{\citenamefont {Wines}\ \emph {et~al.}(2023)\citenamefont {Wines},
  \citenamefont {Tiihonen}, \citenamefont {Saritas}, \citenamefont {Krogel},\
  and\ \citenamefont {Ataca}}]{wines2023quantum}%
  \BibitemOpen
  \bibfield  {author} {\bibinfo {author} {\bibfnamefont {D.}~\bibnamefont
  {Wines}}, \bibinfo {author} {\bibfnamefont {J.}~\bibnamefont {Tiihonen}},
  \bibinfo {author} {\bibfnamefont {K.}~\bibnamefont {Saritas}}, \bibinfo
  {author} {\bibfnamefont {J.~T.}\ \bibnamefont {Krogel}},\ and\ \bibinfo
  {author} {\bibfnamefont {C.}~\bibnamefont {Ataca}},\ }\bibfield  {title}
  {\bibinfo {title} {A quantum monte carlo study of the structural, energetic,
  and magnetic properties of two-dimensional h and t phase vse2},\ }\href
  {https://doi.org/10.1021/acs.jpclett.3c00497} {\bibfield  {journal} {\bibinfo
   {journal} {The Journal of Physical Chemistry Letters}\ }\textbf {\bibinfo
  {volume} {14}},\ \bibinfo {pages} {3553} (\bibinfo {year}
  {2023})}\BibitemShut {NoStop}%
\bibitem [{\citenamefont {{Kolmogorov}}\ and\ \citenamefont
  {{Crespi}}(2005)}]{kolmogorov2005registry}%
  \BibitemOpen
  \bibfield  {author} {\bibinfo {author} {\bibfnamefont {A.~N.}\ \bibnamefont
  {{Kolmogorov}}}\ and\ \bibinfo {author} {\bibfnamefont {V.~H.}\ \bibnamefont
  {{Crespi}}},\ }\bibfield  {title} {\bibinfo {title} {{Registry}-dependent
  interlayer potential for graphitic systems},\ }\href
  {https://journals.aps.org/prb/abstract/10.1103/PhysRevB.71.235415} {\bibfield
   {journal} {\bibinfo  {journal} {{Physical} {Review} B}\ }\textbf {\bibinfo
  {volume} {71}},\ \bibinfo {pages} {235415} (\bibinfo {year}
  {2005})}\BibitemShut {NoStop}%
\bibitem [{\citenamefont {{Ouyang}}\ \emph {et~al.}(2018)\citenamefont
  {{Ouyang}}, \citenamefont {{Mandelli}}, \citenamefont {{Urbakh}},\ and\
  \citenamefont {{Hod}}}]{ouyang2018nanoserpents}%
  \BibitemOpen
  \bibfield  {author} {\bibinfo {author} {\bibfnamefont {W.}~\bibnamefont
  {{Ouyang}}}, \bibinfo {author} {\bibfnamefont {D.}~\bibnamefont
  {{Mandelli}}}, \bibinfo {author} {\bibfnamefont {M.}~\bibnamefont
  {{Urbakh}}},\ and\ \bibinfo {author} {\bibfnamefont {O.}~\bibnamefont
  {{Hod}}},\ }\bibfield  {title} {\bibinfo {title} {{Nanoserpents}: {Graphene}
  nanoribbon motion on two-dimensional hexagonal materials},\ }\href
  {https://pubs.acs.org/doi/10.1021/acs.nanolett.8b02848} {\bibfield  {journal}
  {\bibinfo  {journal} {{Nano} letters}\ }\textbf {\bibinfo {volume} {18}},\
  \bibinfo {pages} {6009} (\bibinfo {year} {2018})}\BibitemShut {NoStop}%
\bibitem [{\citenamefont {{Thompson}}\ \emph {et~al.}(2022)\citenamefont
  {{Thompson}}, \citenamefont {{Aktulga}}, \citenamefont {{Berger}},
  \citenamefont {{Bolintineanu}}, \citenamefont {{Brown}}, \citenamefont
  {{Crozier}}, \citenamefont {in't {Veld}}, \citenamefont {{Kohlmeyer}},
  \citenamefont {{Moore}}, \citenamefont {{Nguyen}} \emph
  {et~al.}}]{thompson2022lammps}%
  \BibitemOpen
  \bibfield  {author} {\bibinfo {author} {\bibfnamefont {A.~P.}\ \bibnamefont
  {{Thompson}}}, \bibinfo {author} {\bibfnamefont {H.~M.}\ \bibnamefont
  {{Aktulga}}}, \bibinfo {author} {\bibfnamefont {R.}~\bibnamefont {{Berger}}},
  \bibinfo {author} {\bibfnamefont {D.~S.}\ \bibnamefont {{Bolintineanu}}},
  \bibinfo {author} {\bibfnamefont {W.~M.}\ \bibnamefont {{Brown}}}, \bibinfo
  {author} {\bibfnamefont {P.~S.}\ \bibnamefont {{Crozier}}}, \bibinfo {author}
  {\bibfnamefont {P.~J.}\ \bibnamefont {in't {Veld}}}, \bibinfo {author}
  {\bibfnamefont {A.}~\bibnamefont {{Kohlmeyer}}}, \bibinfo {author}
  {\bibfnamefont {S.~G.}\ \bibnamefont {{Moore}}}, \bibinfo {author}
  {\bibfnamefont {T.~D.}\ \bibnamefont {{Nguyen}}}, \emph {et~al.},\ }\bibfield
   {title} {\bibinfo {title} {{LAMMPS} - a flexible simulation tool for
  particle-based materials modeling at the atomic, meso, and continuum
  scales},\ }\href
  {https://www.sciencedirect.com/science/article/pii/S0010465521002836}
  {\bibfield  {journal} {\bibinfo  {journal} {{Computer} {Physics}
  {Communications}}\ }\textbf {\bibinfo {volume} {271}},\ \bibinfo {pages}
  {108171} (\bibinfo {year} {2022})}\BibitemShut {NoStop}%
\bibitem [{kro()}]{krongchon2025data}%
  \BibitemOpen
  \href@noop {} {}\bibinfo {note} {See Supplemental Material at
  URL-will-be-inserted-by-publisher for the QMC-fitted LAMMPS potential file
  and the energy data from QMC and the fits.}\BibitemShut {Stop}%
\bibitem [{\citenamefont {{Kim}}\ \emph {et~al.}(2018)\citenamefont {{Kim}},
  \citenamefont {{Baczewski}}, \citenamefont {{Beaudet}}, \citenamefont
  {{Benali}}, \citenamefont {{Bennett}}, \citenamefont {{Berrill}},
  \citenamefont {{Blunt}}, \citenamefont {{Borda}}, \citenamefont {{Casula}},
  \citenamefont {{Ceperley}} \emph {et~al.}}]{kim2018qmcpack}%
  \BibitemOpen
  \bibfield  {author} {\bibinfo {author} {\bibfnamefont {J.}~\bibnamefont
  {{Kim}}}, \bibinfo {author} {\bibfnamefont {A.~D.}\ \bibnamefont
  {{Baczewski}}}, \bibinfo {author} {\bibfnamefont {T.~D.}\ \bibnamefont
  {{Beaudet}}}, \bibinfo {author} {\bibfnamefont {A.}~\bibnamefont {{Benali}}},
  \bibinfo {author} {\bibfnamefont {M.~C.}\ \bibnamefont {{Bennett}}}, \bibinfo
  {author} {\bibfnamefont {M.~A.}\ \bibnamefont {{Berrill}}}, \bibinfo {author}
  {\bibfnamefont {N.~S.}\ \bibnamefont {{Blunt}}}, \bibinfo {author}
  {\bibfnamefont {E.~J.~L.}\ \bibnamefont {{Borda}}}, \bibinfo {author}
  {\bibfnamefont {M.}~\bibnamefont {{Casula}}}, \bibinfo {author}
  {\bibfnamefont {D.~M.}\ \bibnamefont {{Ceperley}}}, \emph {et~al.},\
  }\bibfield  {title} {\bibinfo {title} {{QMCPACK}: an open source \textit{ab
  initio} quantum {Monte} {Carlo} package for the electronic structure of
  atoms, molecules and solids},\ }\href
  {https://iopscience.iop.org/article/10.1088/1361-648X/aab9c3} {\bibfield
  {journal} {\bibinfo  {journal} {{Journal} of {Physics}: {Condensed}
  {Matter}}\ }\textbf {\bibinfo {volume} {30}},\ \bibinfo {pages} {195901}
  (\bibinfo {year} {2018})}\BibitemShut {NoStop}%
\bibitem [{\citenamefont {{Giannozzi}}\ \emph {et~al.}(2009)\citenamefont
  {{Giannozzi}}, \citenamefont {{Baroni}}, \citenamefont {{Bonini}},
  \citenamefont {{Calandra}}, \citenamefont {{Car}}, \citenamefont
  {{Cavazzoni}}, \citenamefont {{Ceresoli}}, \citenamefont {{Chiarotti}},
  \citenamefont {{Cococcioni}}, \citenamefont {{Dabo}} \emph
  {et~al.}}]{giannozzi2009quantum}%
  \BibitemOpen
  \bibfield  {author} {\bibinfo {author} {\bibfnamefont {P.}~\bibnamefont
  {{Giannozzi}}}, \bibinfo {author} {\bibfnamefont {S.}~\bibnamefont
  {{Baroni}}}, \bibinfo {author} {\bibfnamefont {N.}~\bibnamefont {{Bonini}}},
  \bibinfo {author} {\bibfnamefont {M.}~\bibnamefont {{Calandra}}}, \bibinfo
  {author} {\bibfnamefont {R.}~\bibnamefont {{Car}}}, \bibinfo {author}
  {\bibfnamefont {C.}~\bibnamefont {{Cavazzoni}}}, \bibinfo {author}
  {\bibfnamefont {D.}~\bibnamefont {{Ceresoli}}}, \bibinfo {author}
  {\bibfnamefont {G.~L.}\ \bibnamefont {{Chiarotti}}}, \bibinfo {author}
  {\bibfnamefont {M.}~\bibnamefont {{Cococcioni}}}, \bibinfo {author}
  {\bibfnamefont {I.}~\bibnamefont {{Dabo}}}, \emph {et~al.},\ }\bibfield
  {title} {\bibinfo {title} {{QUANTUM ESPRESSO}: a modular and open-source
  software project for quantum simulations of materials},\ }\href
  {https://iopscience.iop.org/article/10.1088/0953-8984/21/39/395502}
  {\bibfield  {journal} {\bibinfo  {journal} {{Journal} of physics: {Condensed}
  matter}\ }\textbf {\bibinfo {volume} {21}},\ \bibinfo {pages} {395502}
  (\bibinfo {year} {2009})}\BibitemShut {NoStop}%
\bibitem [{\citenamefont {{Giannozzi}}\ \emph {et~al.}(2017)\citenamefont
  {{Giannozzi}}, \citenamefont {{Andreussi}}, \citenamefont {{Brumme}},
  \citenamefont {{Bunau}}, \citenamefont {{Nardelli}}, \citenamefont
  {{Calandra}}, \citenamefont {{Car}}, \citenamefont {{Cavazzoni}},
  \citenamefont {{Ceresoli}}, \citenamefont {{Cococcioni}} \emph
  {et~al.}}]{giannozzi2017advanced}%
  \BibitemOpen
  \bibfield  {author} {\bibinfo {author} {\bibfnamefont {P.}~\bibnamefont
  {{Giannozzi}}}, \bibinfo {author} {\bibfnamefont {O.}~\bibnamefont
  {{Andreussi}}}, \bibinfo {author} {\bibfnamefont {T.}~\bibnamefont
  {{Brumme}}}, \bibinfo {author} {\bibfnamefont {O.}~\bibnamefont {{Bunau}}},
  \bibinfo {author} {\bibfnamefont {M.~B.}\ \bibnamefont {{Nardelli}}},
  \bibinfo {author} {\bibfnamefont {M.}~\bibnamefont {{Calandra}}}, \bibinfo
  {author} {\bibfnamefont {R.}~\bibnamefont {{Car}}}, \bibinfo {author}
  {\bibfnamefont {C.}~\bibnamefont {{Cavazzoni}}}, \bibinfo {author}
  {\bibfnamefont {D.}~\bibnamefont {{Ceresoli}}}, \bibinfo {author}
  {\bibfnamefont {M.}~\bibnamefont {{Cococcioni}}}, \emph {et~al.},\ }\bibfield
   {title} {\bibinfo {title} {{Advanced} capabilities for materials modelling
  with {Quantum} {ESPRESSO}},\ }\href
  {https://iopscience.iop.org/article/10.1088/1361-648X/aa8f79} {\bibfield
  {journal} {\bibinfo  {journal} {{Journal} of physics: {Condensed} matter}\
  }\textbf {\bibinfo {volume} {29}},\ \bibinfo {pages} {465901} (\bibinfo
  {year} {2017})}\BibitemShut {NoStop}%
\bibitem [{\citenamefont {{Giannozzi}}\ \emph {et~al.}(2020)\citenamefont
  {{Giannozzi}}, \citenamefont {{Baseggio}}, \citenamefont {{Bonf}{\`a}},
  \citenamefont {{Brunato}}, \citenamefont {{Car}}, \citenamefont {{Carnimeo}},
  \citenamefont {{Cavazzoni}}, \citenamefont {{De}~{Gironcoli}}, \citenamefont
  {{Delugas}}, \citenamefont {{Ferrari}~{Ruffino}} \emph
  {et~al.}}]{giannozzi2020quantum}%
  \BibitemOpen
  \bibfield  {author} {\bibinfo {author} {\bibfnamefont {P.}~\bibnamefont
  {{Giannozzi}}}, \bibinfo {author} {\bibfnamefont {O.}~\bibnamefont
  {{Baseggio}}}, \bibinfo {author} {\bibfnamefont {P.}~\bibnamefont
  {{Bonf}{\`a}}}, \bibinfo {author} {\bibfnamefont {D.}~\bibnamefont
  {{Brunato}}}, \bibinfo {author} {\bibfnamefont {R.}~\bibnamefont {{Car}}},
  \bibinfo {author} {\bibfnamefont {I.}~\bibnamefont {{Carnimeo}}}, \bibinfo
  {author} {\bibfnamefont {C.}~\bibnamefont {{Cavazzoni}}}, \bibinfo {author}
  {\bibfnamefont {S.}~\bibnamefont {{De}~{Gironcoli}}}, \bibinfo {author}
  {\bibfnamefont {P.}~\bibnamefont {{Delugas}}}, \bibinfo {author}
  {\bibfnamefont {F.}~\bibnamefont {{Ferrari}~{Ruffino}}}, \emph {et~al.},\
  }\bibfield  {title} {\bibinfo {title} {{Quantum} {ESPRESSO} toward the
  exascale},\ }\href {https://doi.org/10.1063/5.0005082} {\bibfield  {journal}
  {\bibinfo  {journal} {{The} {Journal} of chemical physics}\ }\textbf
  {\bibinfo {volume} {152}},\ \bibinfo {pages} {154105} (\bibinfo {year}
  {2020})}\BibitemShut {NoStop}%
\bibitem [{\citenamefont {{Bennett}}\ \emph {et~al.}(2017)\citenamefont
  {{Bennett}}, \citenamefont {{Melton}}, \citenamefont {{Annaberdiyev}},
  \citenamefont {{Wang}}, \citenamefont {{Shulenburger}},\ and\ \citenamefont
  {{Mitas}}}]{bennett2017new}%
  \BibitemOpen
  \bibfield  {author} {\bibinfo {author} {\bibfnamefont {M.~C.}\ \bibnamefont
  {{Bennett}}}, \bibinfo {author} {\bibfnamefont {C.~A.}\ \bibnamefont
  {{Melton}}}, \bibinfo {author} {\bibfnamefont {A.}~\bibnamefont
  {{Annaberdiyev}}}, \bibinfo {author} {\bibfnamefont {G.}~\bibnamefont
  {{Wang}}}, \bibinfo {author} {\bibfnamefont {L.}~\bibnamefont
  {{Shulenburger}}},\ and\ \bibinfo {author} {\bibfnamefont {L.}~\bibnamefont
  {{Mitas}}},\ }\bibfield  {title} {\bibinfo {title} {A new generation of
  effective core potentials for correlated calculations},\ }\href
  {https://doi.org/10.1063/1.4995643} {\bibfield  {journal} {\bibinfo
  {journal} {{The} {Journal} of chemical physics}\ }\textbf {\bibinfo {volume}
  {147}} (\bibinfo {year} {2017})}\BibitemShut {NoStop}%
\bibitem [{\citenamefont {{Drummond}}\ \emph {et~al.}(2008)\citenamefont
  {{Drummond}}, \citenamefont {{Needs}}, \citenamefont {{Sorouri}},\ and\
  \citenamefont {{Foulkes}}}]{drummond2008finite}%
  \BibitemOpen
  \bibfield  {author} {\bibinfo {author} {\bibfnamefont {N.}~\bibnamefont
  {{Drummond}}}, \bibinfo {author} {\bibfnamefont {R.}~\bibnamefont {{Needs}}},
  \bibinfo {author} {\bibfnamefont {A.}~\bibnamefont {{Sorouri}}},\ and\
  \bibinfo {author} {\bibfnamefont {W.}~\bibnamefont {{Foulkes}}},\ }\bibfield
  {title} {\bibinfo {title} {{Finite}-size errors in continuum quantum {Monte}
  {Carlo} calculations},\ }\href
  {https://journals.aps.org/prb/abstract/10.1103/PhysRevB.78.125106} {\bibfield
   {journal} {\bibinfo  {journal} {{Physical} {Review} B}\ }\textbf {\bibinfo
  {volume} {78}},\ \bibinfo {pages} {125106} (\bibinfo {year}
  {2008})}\BibitemShut {NoStop}%
\bibitem [{\citenamefont {{Brenner}}\ \emph {et~al.}(2002)\citenamefont
  {{Brenner}}, \citenamefont {{Shenderova}}, \citenamefont {{Harrison}},
  \citenamefont {{Stuart}}, \citenamefont {{Ni}},\ and\ \citenamefont
  {{Sinnott}}}]{brenner2002second}%
  \BibitemOpen
  \bibfield  {author} {\bibinfo {author} {\bibfnamefont {D.~W.}\ \bibnamefont
  {{Brenner}}}, \bibinfo {author} {\bibfnamefont {O.~A.}\ \bibnamefont
  {{Shenderova}}}, \bibinfo {author} {\bibfnamefont {J.~A.}\ \bibnamefont
  {{Harrison}}}, \bibinfo {author} {\bibfnamefont {S.~J.}\ \bibnamefont
  {{Stuart}}}, \bibinfo {author} {\bibfnamefont {B.}~\bibnamefont {{Ni}}},\
  and\ \bibinfo {author} {\bibfnamefont {S.~B.}\ \bibnamefont {{Sinnott}}},\
  }\bibfield  {title} {\bibinfo {title} {A second-generation reactive empirical
  bond order {(REBO)} potential energy expression for hydrocarbons},\ }\href
  {https://iopscience.iop.org/article/10.1088/0953-8984/14/4/312} {\bibfield
  {journal} {\bibinfo  {journal} {{Journal} of {Physics}: {Condensed}
  {Matter}}\ }\textbf {\bibinfo {volume} {14}},\ \bibinfo {pages} {783}
  (\bibinfo {year} {2002})}\BibitemShut {NoStop}%
\bibitem [{\citenamefont {{Los}}\ \emph {et~al.}(2017)\citenamefont {{Los}},
  \citenamefont {{Kroes}}, \citenamefont {{Albe}}, \citenamefont {{Gordillo}},
  \citenamefont {{Katsnelson}},\ and\ \citenamefont
  {{Fasolino}}}]{los2017extended}%
  \BibitemOpen
  \bibfield  {author} {\bibinfo {author} {\bibfnamefont {J.}~\bibnamefont
  {{Los}}}, \bibinfo {author} {\bibfnamefont {J.}~\bibnamefont {{Kroes}}},
  \bibinfo {author} {\bibfnamefont {K.}~\bibnamefont {{Albe}}}, \bibinfo
  {author} {\bibfnamefont {R.}~\bibnamefont {{Gordillo}}}, \bibinfo {author}
  {\bibfnamefont {M.}~\bibnamefont {{Katsnelson}}},\ and\ \bibinfo {author}
  {\bibfnamefont {A.}~\bibnamefont {{Fasolino}}},\ }\bibfield  {title}
  {\bibinfo {title} {{Extended} {Tersoff} potential for boron nitride:
  {Energetics} and elastic properties of pristine and defective h-{BN}},\
  }\href {https://journals.aps.org/prb/abstract/10.1103/PhysRevB.96.184108}
  {\bibfield  {journal} {\bibinfo  {journal} {{Physical} {Review} B}\ }\textbf
  {\bibinfo {volume} {96}},\ \bibinfo {pages} {184108} (\bibinfo {year}
  {2017})}\BibitemShut {NoStop}%
\bibitem [{\citenamefont {Szyniszewski}\ \emph {et~al.}(2025)\citenamefont
  {Szyniszewski}, \citenamefont {Mostaani}, \citenamefont {Knothe},
  \citenamefont {Enaldiev}, \citenamefont {Ferrari}, \citenamefont {Fal'ko},\
  and\ \citenamefont
  {Drummond}}]{szyniszewskiAdhesionReconstructionGraphene2025}%
  \BibitemOpen
  \bibfield  {author} {\bibinfo {author} {\bibfnamefont {M.}~\bibnamefont
  {Szyniszewski}}, \bibinfo {author} {\bibfnamefont {E.}~\bibnamefont
  {Mostaani}}, \bibinfo {author} {\bibfnamefont {A.}~\bibnamefont {Knothe}},
  \bibinfo {author} {\bibfnamefont {V.}~\bibnamefont {Enaldiev}}, \bibinfo
  {author} {\bibfnamefont {A.~C.}\ \bibnamefont {Ferrari}}, \bibinfo {author}
  {\bibfnamefont {V.~I.}\ \bibnamefont {Fal'ko}},\ and\ \bibinfo {author}
  {\bibfnamefont {N.~D.}\ \bibnamefont {Drummond}},\ }\bibfield  {title}
  {\bibinfo {title} {Adhesion and {{Reconstruction}} of
  {{Graphene}}/{{Hexagonal Boron Nitride Heterostructures}}: {{A Quantum Monte
  Carlo Study}}},\ }\href {https://doi.org/10.1021/acsnano.4c10909} {\bibfield
  {journal} {\bibinfo  {journal} {ACS Nano}\ }\textbf {\bibinfo {volume}
  {19}},\ \bibinfo {pages} {6014} (\bibinfo {year} {2025})}\BibitemShut
  {NoStop}%
\end{thebibliography}%

\end{document}